\documentclass[10pt]{article}
\usepackage{geometry}                
\usepackage[parfill]{parskip}
\usepackage{amssymb}

\usepackage{amsmath}

\title{\bf Hierarchy of sum rules for oscillator strengths}

\author{C. V. Sukumar \\{\em Wadham College,}\\{\em University of Oxford, Oxford OX1 3PN, U.K. }
}

\begin{document}
\maketitle

\begin{abstract}
It is shown that the well known sum rules for oscillator strengths for Hydrogen atom can be generalized to a whole class of sum rules. The sum rules have contributions from the discrete and the continuum parts of the spectrum neither of which can be calculated in closed analytical form but can be calculated numerically. The numerical calculations are carried out to check the validity of the sum rules. The procedure for constructing sum rules for general potentials is discussed. Generalizations of Kramers relations and the Virial theorem are discussed.

\end{abstract}

\section{Introduction}

The sums of the squares of the amplitudes of the electric dipole matrix elements weighted by a power of the frequency of the incident radiation provide measures of the response of an atom to radiation incident upon it and have many useful applications (Jackiw 1967, Bethe 1964). In this study we examine the generalization of the sum rules for oscillator strengths to the sums of the squared amplitudes of matrix elements for excitations caused by a perturbing potential weighted by powers of the energy differences between the initial state and the excited states connected to the initial state by the perturbation. This report contains an extension and generalization of the results given by Jackiw (1967) and Bethe (1964) and other research work from the early days of Quantum Mechanics (Kramers 1926).

\medskip

{\bf Classical treatment:}

The energy of an oscillating dipole is related to the energy of a simple harmonic oscillator (SHO) and is given by
\begin{equation}
W= \frac{ M \omega^2 A^2}{2}\ = \ \frac{M \omega^2 D^2}{2 e^2}
\end{equation}
where  A is the amplitude of the oscillation of the SHO, M and $\omega$ are the mass and frequency of the oscillator D is the peak value of the dipole moment and $e$ is the fundamental unit of charge. The rate of energy loss by the oscillating dipole is
\begin{equation}
- \frac{dW}{dt}= \frac{2}{3} \frac{1}{4\pi \epsilon_0 c^3} \Big(\frac{d^2p}{dt^2}\Big)^2 
\end{equation} 
where $p=D \cos\omega t$ is the instantaneous value of the dipole moment. The average over many oscillations yields 
\begin{equation}
  <-dW/dt> = \frac{2}{3} \frac{1}{4\pi\epsilon_0 c^3} \frac{\omega^4 D^2}{2}
\end{equation} 
from which we can extract the damping coefficient
\begin{equation}
 \gamma= -\ \frac{<dW/dt>}{W}\ = \frac{2}{3} \frac{e^2}{4\pi \epsilon_0} \frac {\omega^2}{Mc^3} = \frac{2}{3} \alpha_e\ \frac{\hbar\omega}{Mc^2}\ \omega
\end{equation}
where $\alpha_e$ is the fine structure constant.

\bigskip
 
{\bf Quantum treatment:}

Einstein B coefficient:
 
Using time independent perturbation theory and the Rotating Wave Approximation it may be shown that the Einstein coefficient for absorption from a quantum state $\phi_a$ to a quantum state $\phi_b$ by dipole radiation (Foot 2013) is
\begin{equation}
 B_{ab}\ =\  \frac{4 \pi^2}{3\hbar^2}\ \frac{e^2}{4\pi\ \epsilon_0}\ |{\bf r_{ba}}|^2
 \end{equation}
where ${\bf r_{ba}}=<\phi_b|{\bf r}|\phi_a>$ is the dipole matrix element. The spontaneous emission rate can be inferred using the Einstein relation
\begin{equation}
 A_{ba}\ =\ \frac{\hbar\ \omega_{ba}^3}{\pi^2\ c^3}\ B_{ab}
\end{equation}
and may be compared with $\gamma$ evaluated classically. If the Compton wavelength $\lambda$ is used as the scaling distance then 
\begin{equation}
 \lambda\ = \frac{\hbar}{Mc}\ \ \ \  \hbox{and}\ \ A_{ba}\ =\ \frac{4}{3}\ \alpha_e\ \frac{\hbar\ \omega_{ba}}{Mc^2}\ \omega_{ba}\ \Big[\Big(\frac{|{\bf r_{ba}}|}{\lambda}\Big)^2\Big]
\end{equation}

It can be seen that the classical and Quantum expressions for $\gamma$ correspond if the expression in square brackets has the value $1/2$ which is the case for the quantum SHO that we consider next. 

{\bf 3-d Oscillator}:

Using the spherical harmonics the ground  state and the first excited state of the 3-dimensional oscillator may be given in the form
\begin{align}
\phi_a\ &=\ \Big(\frac{1}{\pi^3\ \lambda^6}\Big)^{\frac{1}{4}}\ {\exp \big(-\frac{r^2}{2\ \lambda^2}\big)} \\
\phi_b\ &=\ \Big(\frac{4}{\pi^3\ \lambda^{10}}\Big)^{\frac{1}{4}}\ r\ {\exp \big(-\frac{r^2}{2\ \lambda^2}\big)}\ \cos \theta
\end{align}
in terms of the scale length $\lambda= \sqrt{\frac{\hbar}{M\omega}}$. The dipole matrix element may be evaluated to give $|{\bf r_{ba}}|\ =\ \frac{\lambda}{{\sqrt 2}}$ yielding the Einstein coefficient
\begin{equation}
A_{ba}\ =\  \frac{2}{3}\ \alpha_e\ \frac{\hbar\omega}{Mc^2}\ \omega
\end{equation}
in exact agreement with the classical result. This exact correspondence between the classical and quantum results had been known from the early days of Quantum Mechanics. We next examine the Hydrogen atom to see whether this correspondence has general validity.

\bigskip

{\bf Hydrogen atom}:

The ground state and the first excited p-state of the Hydrogen atom are
\begin{align}
\phi_a\ &=\ {\sqrt{\frac{1}{\pi a_0^3}}}\ {\exp\big(-\frac{r}{a_0}\big)}\\
\phi_b\ &=\ \frac{1}{8}\ {\sqrt{\frac{1}{\pi a_0^3}}}\ \frac{r}{a_0}\ {\exp\big(-\frac{r}{2\ a_0}\big)}\ {\sqrt 2}\ \cos \theta
\end{align}
where the scale length $a_0$ is the Bohr radius which is related to the Compton wavelength by $a_0=\frac{\lambda}{\alpha_e}$. The dipole matrix element may be evaluated to give $|{\bf r_{ba}}| = a_0\ \sqrt{2^{15}/3^{10}}$. The angular frequency of the radiation is $\omega_{ba}= 3/8\ \alpha_e^2\ Mc^2/ \hbar$ and $\ \omega_{ba}\ a_0/c = 3/8\ \alpha_e$. These relations may be used to give
\begin{equation}
A_{ba}\ =\ \frac{4}{3}\ \alpha_e\ \Big(\frac{3}{8}\ \alpha_e\Big)^2\ \frac{2^{15}}{3^{10}}\ \ \omega_{ba}\ =\ \Big(\frac{4}{9}\ \alpha_e\Big)^4\  \frac{c}{a_0}\ =\ \Big(\frac{4}{9}\Big)^4\ \alpha_e^3\  \frac{c}{\lambda} 
\end{equation}
from which the lifetime of the 2p state can be calculated to be approximately 1.6 nanoseconds. The ratio $a_0/c$ is the time taken by light to travel a distance $a_0$. The last form of the expression in terms of the Compton wavelength exhibits the correct cubic dependence on $\alpha_e$ where two powers of $\alpha_e$ arise from the dependence of the energy levels on the electromagnetic interaction and a further power of $\alpha_e$ arises from the electromagnetic interaction responsible for the emission of the photon.

It is clear that the oscillator strengths provide a vital link between theory and the experimental study of the response of a quantum system to radiation incident upon it.

\section{Dipole Sum rules}

For a Hamiltonian with a potential which depends only on the position with eigenstates denoted by $|m\rangle$, where the index $m$ can take values which span both the discrete and continuous spectrum,
\begin{align}
H_0\ &=\ \frac{p^2}{2M}\ +\ V(r)\ \ ,\ \ \ [z,H_0]\ = \frac{1}{2M}\ [z,p^2]\ =\ \frac{i\hbar}{M}\ p_z  \label{eq:I1}\\
\langle m\ |\ [z,H_0]\ |n\rangle\ &=\ (E_{n}\ -E_{m})\ \langle m|\ z\ |n\rangle\ =\ \frac{i\hbar}{M}\ \langle m|\ p_z\ |n\rangle \label{eq:I2}\\
\langle m|\ p_z\ |n\rangle\ &=\ i\ \frac{M}{\hbar}\ (E_m\ -\ E_n)\ \langle m|\ z\ |n\rangle\ =\ i\ M \omega_{mn}\ \langle m|\ z\ |n\rangle \label{eq:I3}
\end{align}
It is evident that the matrix elements of a perturbing potential $V_1$  taken between the eigenstates of $H_0$ satisfy the sum rule
\begin{equation}
\sum_n \ |\langle m|\ V_1\ |n\rangle|^2\ =\ \sum_n \ \langle m|\ V_1\ |n\rangle\ \langle n |\ V_1\ |m\rangle\ =\ \langle m|\ V_1^2\ |m\rangle \label{eq:I4}
\end{equation}
where we have utilized the completeness relation satisfied by the full set of eigenstates of $H_0$ to eliminate the sum over the discrete and continuum eigenstates. A sum rule for the dipole matrix elements may be derived by choosing $V_1 = {\bf r}$ which leads to
\begin{equation}
\sum_n \ |\langle m|\ {\bf r}\ |n\rangle|^2\ =\ \langle m|\ r^2\ |m\rangle \label{eq:I5}
\end{equation}
Continuing this procedure, (\ref{eq:I3}) can be used to establish that
\begin{align}
[z,p_z]\ &=\ i\hbar\ \rightarrow\ \langle m |\ i\hbar\ |m \rangle\ =\ \sum_{n} \Big( \langle m |\ z\ |n\rangle\  \langle n|\ p_z\ |m\rangle\ -\ \langle m|\ p_z\ | n \rangle\ \langle n|\ z\ |m \rangle\Big) \label{eq:I6}\\
i\hbar\ &=\ iM\sum_{n}\ \big( \omega_{nm}\ -\omega_{mn}\big)\ | \langle m|\ z\ |n\rangle |^2  \label{eq:I7}\\
\frac{\hbar^2}{2M}\ &=\ \sum_n \big( E_n\ -\ E_m\big)\ | \langle m|\ z\ |n \rangle |^2 \label{eq:I8}
\end{align}
which is an energy weighted sum rule for the dipole oscillator strengths - this is the famous Thomas-Reiche-Kuhn oscillator strength sum rule (Jackiw 1967) which played an important role in Quantum Mechanics. In equations (\ref{eq:I4}) - (\ref{eq:I8}) for the sake of compactness the summation symbol is used to denote both the sum over the discrete states and the integral over continuum states. We can make this explicit by defining the sums
\begin{equation}
S_j\ =\ \sum_n\ \big(k_m^2-k_n^2\big)^J\  | \langle m|\ z\ |n \rangle |^2\ +\int_0^{\infty}dk\ \big(k_m^2 +k^2\big)^J\ | \langle m|\ z\ |k \rangle |^2 \label{eq:I9}
\end{equation}
where $k_n^2=-2M E_n/\hbar^2$ and $k^2 = 2ME/\hbar^2$ are the wave numbers associated with the discrete and continuum states. (\ref{eq:I5}) can be given in the form $S_0 = \langle m|\ r^2\ |m\rangle$. The energy weighted sum rule (\ref{eq:I8}) becomes 
\begin{equation}
 S_1\ =\ \sum_n\ \big(k_m^2-k_n^2\big)\  | \langle m|\ z\ |n \rangle |^2\ +\int_0^{\infty}dk\ \big(k_m^2 +k^2\big)\ | \langle m|\ z\ |k \rangle |^2\ =\ 1 \label{eq:I10}
\end{equation}

The next higher order sum rule weighted by the squares of the energy differences may be found by using (\ref{eq:I2}) and (\ref{eq:I3}) resulting in the relation 
\begin{equation}
S_2=\sum_n \big(k_m^2-k_n^2\big)^2  | \langle m| z |n \rangle |^2 +\int_0^{\infty}dk \big(k_m^2 +k^2\big)^2 | \langle m| z |k \rangle |^2 = \frac{4}{\hbar^2}  \langle m| p_z^2 |m \rangle = \frac{4M}{\hbar^2} \langle m| \frac{\partial V_0}{\partial z} z\ |m \rangle  \label{eq:I11}
\end{equation}
where the last equality is valid for the eigenstates of a spherically symmetric potentials and follows from the Virial theorem which provides a relation between the expectation value of twice the kinetic energy and the expectation value of $-\nabla V_0.{\bf r}$ .

The procedure adopted to derive the sum rules may be continued to derive sum rules with higher orders of energy weightings. This can be achieved by finding relations analogous to (\ref{eq:I3}).
For example,
\begin{equation}
\big(E_n - E_m\big)^2\ \langle m|\ z\ |n \rangle \ =\ \langle m|\ [[z,H],H]\ |n \rangle \ =\ \frac{i\hbar}{M}\ \langle m|\ [p_z,H]\ |n \rangle\ =\ \frac{\hbar^2}{M}\ \langle m|\ \frac{\partial V_0}{\partial z}\ |n \rangle \label{eq:I12}
\end{equation}  
and this relation may be used together with (\ref{eq:I3}) to show that
\begin{equation}
S_3\ =\ \sum_n\ \big(k_m^2-k_n^2\big)^3\  | \langle m|\ z\ |n \rangle |^2\ +\int_0^{\infty}dk\ \big(k_m^2 +k^2\big)^3\ | \langle m|\ z\ |k \rangle |^2\ =\ -8i\ \frac{M}{\hbar^3}\  \langle m|\ \frac{\partial V_0}{\partial z}\ p_z\ |m \rangle\   
\end{equation} \label{eq:I13}
If the relation
\begin{equation}
-\ \Big\langle \frac{\partial V_0}{\partial z}\ \frac{\partial}{\partial z}\Big\rangle\ =\  \Big\langle \frac{\partial}{\partial z}\frac{\partial V_0}{\partial z}\ \Big\rangle\ =\ \frac{1}{2}\ \Big\langle \Big[\frac{\partial}{\partial z},\frac{\partial V_0}{\partial z}\Big]\ \Big\rangle\ =\ \frac{1}{2}\ \Big\langle \Big(\frac{\partial^2 V_0}{\partial z^2}\Big)\ \Big\rangle \label{eq:I14}
\end{equation} 
is valid then the sum rule may be expressed in the form
\begin{equation}
S_3\ =\ \sum_n\ \big(k_m^2-k_n^2\big)^3\  | \langle m|\ z\ |n \rangle |^2\ +\int_0^{\infty}dk\ \big(k_m^2 +k^2\big)^3\ | \langle m|\ z\ |k \rangle |^2\ =\ 4\ \frac{M}{\hbar^2}\  \langle m|\ \frac{\partial^2 V_0}{\partial z^2}\ |m \rangle \label{eq:I15}   
\end{equation}
which for $l=0$ states in a spherically symmetric potential $V_0$ further simplifies to
\begin{equation}
S_3\ =\ \sum_n\ \big(k_m^2-k_n^2\big)^3\  | \langle m|\ z\ |n \rangle |^2\ +\int_0^{\infty}dk\ \big(k_m^2 +k^2\big)^3\ | \langle m|\ z\ |k \rangle |^2\ =\ \frac{4}{3}\ \frac{M}{\hbar^2}\  \langle m|\ \nabla^2V_0\ |m \rangle \label{eq:I16}   
\end{equation}
We shall return later to the study of the circumstances under which (\ref{eq:I14}) is valid.
  
Similarly (\ref{eq:I12}) may be used twice to show that the fourth order sum rule is of the form
\begin{equation}
S_4=\sum_n\ \big(k_m^2-k_n^2\big)^4\  | \langle m|\ z\ |n \rangle |^2\ +\int_0^{\infty}dk\ \big(k_m^2 +k^2\big)^4\ | \langle m|\ z\ |k \rangle |^2\ =\ 16\ \frac{M^2}{\hbar^4}\  \langle m|\ \frac{\partial V_0}{\partial z}\ \frac{\partial V_0}{\partial z}\ |m \rangle  \label{eq:I17}  
\end{equation}

Proceeding along these lines we can establish a hierarchy of sum rules (Jackiw 1967). However the sum rules are only meaningful when the two sides of the equality lead to finite values. The number of sum rules which lead to finite sums will depend on the potential and the angular momentum of the initial state. The sum rules for Hydrogen can be used to illustrate how a finite number of useful sum rules arise for a specified angular momentum of the init1al state in a Coulomb potential.

{\bf Sum rules for the 1S state of Hydrogen}

The Coulomb potential of an electron in the field of a proton can be given in the equivalent forms
\begin{equation}
V_0(r) \ =\ - \frac{\alpha_e c \hbar}{r}\ =\ -\ \alpha_e^2\ Mc^2\ \frac{a_0}{r}\ =\ -\frac{\hbar^2}{M a_0^2}\ \frac{a_0}{r}\ =\ -\frac{\hbar^2}{M a_0^2}\ \frac{1}{\rho} \label{eq:I18}
\end{equation}
using the Bohr radius $a_0$ as the scaling length. The eigenstates of the electron will be denoted by $|nlm\rangle$. We first examine the sum rules arising from the dipole operator $r\cos \theta = a_0\ z$, expressed in terms of a dimensionless $z$, which can cause excitation from $|100\rangle$ to the states $|n10\rangle$ with $n>1$. Since $E_n= -  \alpha_e^2 Mc^2/(2n^2)$ for the discrete states and $E_k\ = \hbar^2 k^2/(2M) =  \alpha_e^2 Mc^2 q^2/2$, where $q=k a_0$ is dimensionless, we consider the dimensionless quantities 
\begin{equation}
S_J^+ = \big(a_0^2\big)^{J-1}\ S_J\ = \ \sum_{n=2}^{\infty} \Big(\frac{n^2-1}{n^2}\Big)^J\ |\langle 100|\ z\ |n10\rangle |^2\ +\ \int_0^{\infty}dq\  \big(1+q^2\big)^J\ |\langle 100|\ z\ |q00\rangle |^2 \label{eq:I19}
\end{equation}
which are simply related to the sums expressed in terms of $k_n$ and $k$ considered in (\ref{eq:I9}) earlier by a multiplicative factor. The dipole matrix elements can be explicitly evaluated to be
\begin{align}
| \langle 100| z |n00 \rangle |^2\ &=\ \frac{1}{3}\ 2^8\ \frac{n^7}{(n^2-1)^5}\ \frac{(n-1)^{2n}}{(n+1)^{2n}} \label{eq:I20}\\
| \langle 100| z |q00 \rangle |^2\ &=\ \frac{1}{3}\ 2^8\ \frac{q}{(1+q^2)^5}\ \exp\Big(-4\ \frac{\tan^{-1}q}{q}\Big)\ \frac{1}{\big(1 - \exp\big(-\frac{2\pi}{q}\big)\big)} \label{eq:I21}
\end{align}
and the integration variable may be changed by the substitution $q\rightarrow \tan u$ to express $S_J$ in the form
\begin{equation}
S_J^+ = \frac{2^8}{3}\ \Big(\sum_{n=2}^{\infty} \Big( n^{7-2J}\ \frac{(n-1)^{2n+J-5}}{(n+1)^{2n-J+5}} + \int_0^{\frac{\pi}{2}} du \sin u\ (\cos u)^{7-2J} \frac{ \exp(-4 u \cot u)}{\big(1 - \exp\big(-{2\pi} \cot u\big)\big)}\Big) \label{eq:I22}
\end{equation}

According to the sum rules derived earlier we should find that 
\begin{align}
S_0^+\ &=\ \langle100\ |\ z^2\ |\ 100\rangle\ =\ \frac{1}{3}\ \langle100\ |\ \rho^2\ |\ 100\rangle\ =\ 1  \label{eq:J1}\\
S_1^+\ &=\ \langle100\ |\ 100\rangle\ =\ 1  \label{eq:J2}\\
S_2^+\ &=\ 4\ \langle100\ |\ \frac{\partial V_0}{\partial z}\ z\ |\ 100\rangle\ =\ 4\ \langle100\ |\ \frac{\cos^2\theta}{\rho}\ \rho\ |\ 100\rangle\ =\ \frac{4}{3} \label{eq:J3}\\
S_3^+\ &=\  \frac{4}{3}\ \langle 100\ |\ \nabla^2 V_0\ |\ 100\rangle\ =\ \frac{16}{3} \label{eq:J4}
\end{align}
by explicit evaluation of right hand side of the sum rules for the $1S$ state. It is evident from (\ref{eq:I22}) that the continuum contribution to the sum rule diverges for $J>3$ since the integrand develops a divergence at $u=\pi/2$ for $J>3$. The contribution to the sum rules from the discrete and continuum states may be calculated numerically and are given below.
\begin{align}
Order &\ \ \ Discrete && Continuum &&& Sum \ \ \ \  \notag      \\
 S_0^+       &\ \ \  0.716587..  &&\ \ \  0.283412..    &&&\ \ \  0.99999..\\
 S_1^+       &\ \ \  0.565003..  &&\ \ \  0.434996..    &&&\ \ \  0.99999..\\
 S_2^+       &\ \ \  0.449355..  &&\ \ \  0.883977..    &&&\ \ \  1.33333..\\
 S_3^+       &\ \ \  0.360841..  &&\ \ \  4.972492..    &&&\ \ \  5.33333..
 \end{align}
verifying that the addition of the discrete sums which included the first $2000$ terms and the numerical integrations of the continuum contributions indeed converge towards the values stipulated by the sum rules.

It is interesting to note that both the discrete and continuum contributions to $S_J$ can be viewed as arising from integrals of the same integrand integrated along different contours. To show this we first note that
\begin{equation}
\exp\Big(-4n\ \tanh^{-1}\Big(\frac{1}{n}\Big)\Big)\ =\ \exp\Big(-2n\ \ln\Big(\frac{n+1}{n-1}\Big)\Big)\ =\ \Big(\frac{n-1}{n+1}\Big)^{2n}
\end{equation}
and treat $n$ as a continuous variable and transform to a new variable $v=n^{-1}$ so that $dn= -v^{-2}\ dv$. These manipulations lead us to consider the integral
\begin{equation}
I_J\ =\ -\frac{2^8}{3}\ \int_{C} dv\ v\ {\big(1-v^2\big)^{J-5}}\ \exp\Big(-4\ \frac{\tanh^{-1}v}{v}\Big)\ \frac{1}{\big(1 - \exp\big(-\frac{2\pi i}{v}\big)\big)}
\end{equation}   
which has poles at $v=n^{-1}$ for integer values of $n$ in the complex $v$ plane. If the contour $C$ is chosen as a closed path $C_1$ surrounding all the poles along the positive real $v$ axis except the pole at $v=1$ then the contribution from the discrete states to $S_J$ is equal to the sum over the residues at the poles of the integrand and the continuum contribution to $S_J$ is equal to the contribution arising from a line integral of the same integrand taken along the contour $C_2$ which is the positive imaginary $v$ axis. 

\section{ Constructive Approach to Sum rules}
A constructive approach to the study of sum rules of all orders would be to consider the hierarchy of functions defined by
\begin{align}
F_0\ &=\ V_1\ |m\rangle \label{eq:I23}\\
F_{K+1}\ &=\ \Big(\frac{2M a^2}{\hbar^2}\Big)\ \big( H_0\ -\ E_m\big) \ F_K\ \ ,\ K=0,1,2....\label{eq:I24}
\end{align}
which can all be calculated directly if $H_0$ and $V_1$ are given. We associate a scaling length $a$ with the potential $V_0$ so that dimensionless operators can be used to construct the sum rules. We note that $\langle m\ |\ V_1\ \big(H_0-E_m \big)^J\ V_1\ |\ m\rangle $ will give rise to a sum rule of order $J$ when a complete set of states is inserted anywhere between the $J$ factors of $(H_0-E_m)$ which lie between the the first and second $V_1$. There are $(J+1)$ different ways of doing this. In terms of $F_K$ and the wave numbers the sum rules can be expressed in the form
\begin{align}
S_0\ &=\ \sum_n |\langle m|\ V_1\ |n\rangle |^2\ +\ \int_0^{\infty}dk |\langle m|\ V_1\ | k\rangle |^2\ =\ \langle F_0 | F_0\rangle \label{eq:I25}\\
S_1\ &=\ \sum_n (k_m^2-k_n^2)\ |\langle m|\ V_1\ |n\rangle |^2\ +\ \int_0^{\infty}dk\  (k_m^2+k^2)\ |\langle m|\ V_1\ | k\rangle |^2\ =\ \langle F_0 | F_1\rangle \label{eq:I26}\\
S_2\ &=\ \sum_n (k_m^2-k_n^2)^2\ |\langle m|\ V_1\ |n\rangle |^2\ +\ \int_0^{\infty}dk\  (k_m^2+k^2)^2\ |\langle m|\ V_1\ | k\rangle |^2\ =\ \langle F_1 | F_1\rangle \label{eq:I27}\\
S_3\ &=\ \sum_n (k_m^2-k_n^2)^3\ |\langle m|\ V_1\ |n\rangle |^2\ +\ \int_0^{\infty}dk\  (k_m^2+k^2)^3\ |\langle m|\ V_1\ | k\rangle |^2\ =\ \langle F_1 | F_2\rangle \label{eq:I28}
\end{align}
with obvious generalization to higher orders. It is also evident that the sum rule of order $J$ may be expressed in terms of the overlap of any combination of $F_K$ and $F_N$ as long as $K+N=J$. 

The scheme outlined above may be illustrated with the example of the Hydrogen atom when the state $|m\rangle$ is the $1S$ state. For the dipole excitations it is appropriate to consider the Hamiltonian for the radial Schr{\"{o}}dinger equation partial wave $l=1$ and the solutions $F_K$ can be factorized in the form 
\begin{equation}
F_K\ =\ \frac{1}{{\sqrt 3}}\ Y_{10}\ {\tilde F}_K^+  \label{eq:I29}
\end{equation}
where $Y_{10}$ is a spherical harmonic. In terms of the dimensionless variable $\rho= {r}/{a_0}$,
\begin{equation}
\Big(\frac{2M a^2}{\hbar^2}\Big)\ (H_0\ -\ E_1)\ =\ \Big(-\frac{\partial^2}{\partial\rho^2}\ +\ \frac{2}{\rho^2}\ -\ \frac{2}{\rho}\ +\ 1\ \Big) \label{eq:I30}
\end{equation}
and the first few members of the sequence of ${\tilde F}_m^+$ are
\begin{align}
{\tilde F}_0^+\ &=\ 2\ \rho^2\ \exp(-\rho) \label{eq:I31}\\
{\tilde F}_1^+\ &=\ 4\ \rho\ \exp(-\rho) \label{eq:I32}\\
{\tilde F}_2^+\ &=\ \frac{8}{\rho}\ \exp(-\rho) \label{eq:I33}
\end{align}
The sums $S_J$ considered earlier may now be expressed in terms of $F_m$ as
\begin{align}
S_0^+\ &=\ \ \langle F_0 | F_0\rangle\ =\ \frac{1}{3}\ 2^2\ \frac{4!}{2^5}\ =\ 1 \\
S_1^+\ &=\ \langle F_0 | F_1\rangle\ =\ \frac{1}{3}\ 2^3\ \frac{3!}{2^4}\ =\ 1 \\ 
S_2^+\ &=\ \langle F_1 | F_1\rangle\ =\ \frac{1}{3}\ 2^4\ \frac{2!}{2^3}\ =\ \frac{4}{3} \\
S_3^+\ &=\ \langle F_1 | F_2\rangle\ =\ \frac{1}{3}\ 2^5\ \frac{0!}{2}\ =\ \frac{16}{3}
\end{align}  
in agreement with (\ref{eq:J1}) - (\ref{eq:J4}), verifying the efficacy of the constructive approach to the evaluation of the sum rules. The constructive approach provides a general method for generating the sequence of functions $F_K$ and the evaluation of the sum rules reduces to the evaluation of the overlap integrals of $F_K$.

Until now we have examined sums weighted by positive powers of the energy differences. But the general approach outlined in this section suggests that the procedure may be generalized to negative values of $J$. This will be examined next.

\subsection{Sum rules of negative order}

It is possible to study sums of squares of matrix elements weighted by negative powers of the energy differences by extending the system of equations (\ref{eq:I23}) and (\ref{eq:I24}) to negative values of $m$. For simplicity we first consider the case such as the ground state $1S$ of Hydrogen, when the operator $H_0$ has bound state eigenfunctions for angular momentum $l$ and energy $E_{m}$ but does not have normalizable solutions of angular momenta $L\pm1$ at energy $E_{m}$ and will return the more general case later. For the simpler case, if we define $G_J = F_{-J}$, the  appropriate system of differential equations  is
\begin{align}
G_0\ =\ F_0\ &=\ V_1 |m \rangle  \label{eq:I34}\\
\Big(\frac{2M a^2}{\hbar^2}\Big)\ \big( H_0\ -\ E_m\big) \ G_{K+1}\ &= \ G_K \ \ ,\ \ K=0,1,2,... \label{eq:I35}
\end{align}
which may be solved. In general differential equations with source terms present on the right hand side can have arbitrary amounts of solutions of the homogeneous differential equation added to any solution of the inhomogeneous differential equation and the boundary conditions satisfied at the origin and in the asymptotic region depending upon $V_1$ and the boundary conditions satisfied by $G_0$ need to be taken into account in choosing an appropriate $G_K$. However for the simpler case, such as the ground state of Hydrogen, such complications do not arise. In terms of the solutions to (\ref{eq:I34}) and (\ref{eq:I35}) the sum rules of negative order become
\begin{equation}
S_{-J}\ =\ \sum_{n\ne m}\ \frac{|\langle m| V_1 |n\rangle |^2}{(k_m^2-k_n^2)^{J}}\  +\ \int_0^{\infty}dk\  \frac{|\langle (m| V_1 | k\rangle |^2}{(k_m^2+k^2)^{J}}\   =\ \langle G_K | G_{J-K}\rangle \ \ ,\ K=0,1,..,J \label{eq:I36}
\end{equation}
where we have explicitly indicated $(J+1)$ representations of the sum rule not all of which are distinct because the overlap of $G_K$ and $G_N$ is the same as the overlap of $G_N$ and $G_K$ for real functions.

We illustrate the procedure for the ground state of Hydrogen.  The solutions $G_K$ satisfy the same boundary conditions as the $p$ state bound states of $H_0$. The solutions $G_K$ can be explicitly solved for this case and the first few solutions are
\begin{align}
G_0\ &=\ \frac{2}{{\sqrt 3}}\ Y_{10}\ \rho^2\ \exp(-\rho) \label{eq:I37}\\
G_1\ &=\ G_0\ \frac{1}{2}\  \big(1 + \frac{\rho}{2}\big) \label{eq:I38}\\
G_2\ &=\ G_0\ \frac{1}{48}\ \big(22 + 11\ \rho + 2 \rho^2\big) \label{eq:I39}
\end{align}
which lead to the sum rules
\begin{align}
S_{-1}^+ &= \langle G_0 | G_1\rangle  = \frac{1}{2}\ \frac{1}{3}\ 2^2\ \Big(\frac{4!}{2^5}\ +\ \frac{1}{2}\ \frac{5!}{2^6}\Big) = \frac{9}{8}  = 1.125 \\
S_{-2}^+ &= \langle G_1 | G_1\rangle = \frac{1}{2^2}\ \frac{1}{3}\ 2^2\ \Big(\frac{4!}{2^5}\ +\ \frac{5!}{2^6}\ +\ \frac{1}{4}\ \frac{6!}{2^7}\Big) =  \frac{43}{32} = 1.34375\\ 
S_{-3}^+ &= \langle G_1 | G_2\rangle = \frac{1}{2^3}\ \frac{1}{3}\ 2^2\ \Big(\frac{11}{6} \frac{4!}{2^5}\ +\ \frac{11}{6} \frac{5!}{2^6}\ +\ \frac{5}{8}\ \frac{6!}{2^7} + \frac{1}{12}\ \frac{7!}{2^8}\Big) = \frac{319}{192} = 1.661458.. \\
S_{-4}^+ &= \langle G_2 | G_2\rangle = \frac{1}{2^4} \frac{1}{3} \frac{2^2}{12^2} \Big(484 \frac{4!}{2^5} + 484 \frac{5!}{2^6} + 209 \frac{6!}{2^7} + 44 \frac{7!}{2^8} + 4 \frac{8!}{2^9} \Big) = \frac{9673}{4608} = 2.099175..
\end{align}
The contribution of the discrete and continuum states to the sum rules can be calculated numerically and are given below:
\begin{align}
Order &\ \ \ Discrete && Continuum &&& Sum\ \ \ \   \notag      \\
 S_{-1}^+       &\ \ \  0.915814..  &&\ \ \  0.209185..    &&&\ \ \  1.124999..\label{eq:I40}\\
 S_{-2}^+       &\ \ \  1.178262..  &&\ \ \  0.165487..    &&&\ \ \  1.343749..\\
 S_{-3}^+       &\ \ \  1.524670..  &&\ \ \  0.136787..    &&&\ \ \  1.661457..\\
 S_{-4}^+       &\ \ \  1.982648..  &&\ \ \  0.116526..    &&&\ \ \  2.099174..
\end{align}
verifying that the addition of the discrete sums which included the first $2000$ terms and the numerical integrations of the continuum contributions indeed converge towards the values stipulated by the sum rules. 

We note that in particular the sum rule of order -1 can be related to the polarizability of the ground state of Hydrogen. The second order perturbation shift of the ground state energy of a Hydrogen atom when placed in a uniform electric field polarized along the $z$ direction is
\begin{equation}
W_2\ =\ -\frac{2 M}{\hbar^2}\ ( \alpha c \hbar )\ \Big(\ \sum_{n=2}^{\infty}\ \frac{|\langle\ 1|\ \epsilon z\ |n\ \rangle |^2}{(k_1^2\ -\ k_n^2)}\  +\ \int_0^{\infty}dk\  \frac{|\langle\ 1|\ \epsilon z\ |\ k\ \rangle |^2}{(k_1^2\ +\ k^2)}\ \Big) 
\end{equation}
is related to $S_{-1}$ by
\begin{equation}
S_{-1}^{+}\ =\ \Big(\frac{\hbar^2}{2M a_0^2}\Big)\ \frac{1}{a_0^2}\ \frac{1}{\alpha c \hbar}\ \Big(\frac{-W_2}{\epsilon^2}\Big)\ =\ -\ \frac{1}{2 a_0^3}\ \frac{W_2}{\epsilon^2}\ =\ \frac{9}{8}
\end{equation}  
from which the polarizability of the $1S$ state of Hydrogen can be calculated in leading order (Schiff 1968)
to be
\begin{equation}
\alpha_0\ =\ \frac{9}{2}\ a_0^3
\end{equation}
It is evident from (\ref{eq:I40}) that the bound states provide by far the greater contribution to the ground state polarizability but the continuum contribution is not insignificant.

\subsection{Green's function for negative order sum rules for Coulomb potential}

For the case of the Coulomb potential it is possible to develop a Green's function approach to solve the system of equations (\ref{eq:I34}) and (\ref{eq:I35}). We note that eq.(\ref{eq:I30}) has a factorisable representation which may be used to find solutions to
\begin{equation}
\big(H_0\ -\ E_1\big)\ \Phi \ =\ \frac{-\hbar^2}{2 M a_0^2}\ \Big( \frac{\partial}{\partial \rho}\ +\ 1 \Big)\ \Big( \frac{\partial}{\partial \rho}\ +\ \frac{2}{\rho}\ -1 \Big)\ \Phi\ =\ 0
\end{equation}
Two solutions to this equation one of which vanishes in the asymptotic region and the other is regular at the origin and their Wronskian are
\begin{align}
\Phi_1\ &=\ \exp(-\rho)\ \Big( 1\ +\ \frac{1}{\rho}\ +\ \frac{1}{2\rho^2} \Big) \\
\Phi_2\ &=\ \frac{1}{2 \rho^2}\ \exp(\rho)\ - \ \Phi_1 \\
W_{12}\ &=\ \Phi_2\ \frac{\partial \Phi_1}{\partial\rho}\ -\ \Phi_1\ \frac{\partial \Phi_2}{\partial \rho}\ =\ -\frac{1}{\rho^2} 
\end{align}
These solutions may be used to construct a Green's function which solves
\begin{equation}
\Big( \frac{\partial}{\partial \rho}\ +\ 1 \Big)\ \Big( \frac{\partial}{\partial \rho}\ +\ \frac{2}{\rho}\ -1 \Big)\ G(\rho ,{\tilde \rho})\ =\ W_{12}\ \delta\big(\rho\ -\ {\tilde \rho}\big)
\end{equation}
in terms of which the solution to eq.(\ref{eq:I35}) may be given in the form
\begin{align}
\frac{\hbar^2}{2Ma_0^2}\ G_{K+1}(\rho)\ &=\ \int_0^{\infty}\ G_K({\tilde \rho})\ {\tilde \rho}^2\  G(\rho ,{\tilde \rho})\ d{\tilde \rho} \notag \\
\ &=\  \Phi_1\big(\rho\big)\ \int_0^{\rho} \Phi_2\big({\tilde \rho}\big)\ G_K\big({\tilde \rho}\big)\ {\tilde \rho}^2\ d{\tilde \rho}\ +\ \Phi_2\big(\rho\big)\ \int_{\rho}^{\infty} \Phi_1\big({\tilde \rho}\big)\ G_K\big({\tilde \rho}\big)\ {\tilde \rho}^2\ d{\tilde \rho} 
\end{align}
Using this hierarchy of solutions it is possible to construct sum rules of any negative order. It is evident from (\ref{eq:I22}) that both the discrete sum and the integral converge for all negative $J$. It is also clear that in the limit of large negative $J$ the ratio of successive sum rules tends to $4/3$ as the $n=2$ term from the discrete sum dominates in this limit.

Until now we have considered sum rules for dipole matrix elements connecting the ground state to the $p$ states in the discrete and continuous spectrum. If the initial state is not the ground state and if the initial angular momentum $l\ne 0$ then the procedure we have outlined would need to be modified. We examine this next.

\section{Generalization to arbitrary states}

If the angular momentum of the initial state $l\ne 0$ then the perturbation $V_1$ can cause excitations to states with $l\rightarrow l+1$ and $l\rightarrow l-1$ and if there are degeneracies present, as in the case of Hydrogen $2S$ and $2P$, for example, then there would be normalizable solutions to $(H_0 -E_{m}) =0$ for angular momenta $l,(l\pm 1$) for the same $E_m$ and under these circumstances (\ref{eq:I23}),(\ref{eq:I24}), (\ref{eq:I34}) and (\ref{eq:I35}) would need to be considered with due care. Towards this end it is convenient to separate out the angular part of the relevant equations and examine the radial part expressed in terms of a dimensionless radial coordinate $\rho$ defined by $r=\rho a$ where $a$ is the scaling length associated with the potential $V_0$. Let 
\begin{align}
h_0\ &=\ \frac{2Ma^2}{\hbar^2} \ H_0 \ =\ -\ \frac{\partial^2}{\partial \rho^2}\ +\ \frac{{\bf L}^2}{\rho^2}\ +\ 2 v_0\ \ \ ,\ \ \ v_0\ =\ \frac{Ma^2}{\hbar^2}\ V_0 \\
{\bf L}^2\ P_l(\cos  \theta)\ &=\ l(l+1)\ P_l(\cos \theta)
\end{align}
and the initial state expressed in terms of the normalized radial eigenfunctions $R_{m l}$ be
\begin{equation}
|m,l \rangle\ =\ R_{ml}(\rho)\ \sqrt{\frac{2l+1}{2}}\ P_l(\cos\theta) \ \ {\hbox {and}}\ \ 
(h_0 + k_m^2)\ |m, l \rangle\ =\ 0\ . \label{eq:I41}
\end{equation}
In the above equations and in subsequent discussions $z$ and $k_m$ are dimensionless quantities arising after an appropriate scaling. If the scaled perturbation is $v_1=\rho \cos\theta$, then 
\begin{align}
\cos\theta \ P_l(\cos\theta) &= \frac{l+1}{2l+1}\ P_{l+1}(\cos\theta)\ +\ \frac{l}{2l+1}\ P_{l-1}(\cos\theta)\\
v_1 |m, l\rangle &= \rho R_{m l} \Big( \alpha \sqrt{\frac{2l+3}{2}} P_{l+1} +\beta \sqrt{\frac{2l-1}{2}}\ P_{l-1} \Big) = \rho \Big(\alpha |m, l+1\rangle + \beta |m ,l-1\rangle\Big) \label{eq:K50}
\end{align}
where
\begin{equation}
\alpha\ =\ \frac{l+1}{\sqrt{(2l+1)(2l+3)}}\ \ ,\ \ \beta\ =\ \frac{l}{\sqrt{(2l+1)(2l-1)}} \label{eq:K0}
\end{equation}
and
\begin{align}
\alpha^2\ +\ \beta^2\ &=\ <l| \cos^2 \theta |l>\ =\ \frac{2l(l+1)-1}{4l(l+1)-3} \label{eq:K51}\\
(l+1)\ \alpha^2\ -\ l\ \beta^2\ &=\ \frac{1}{2}\ <m | \sin^2 \theta | m> \ = \ \frac{l(l+1)-1}{4l(l+1)-3}
\end{align}
The orthogonality of Legendre polynomials ensures that for $l>0$ there will be separate sum rules for transitions $l\rightarrow l+1$ and $l\rightarrow l-1$. To exhibit this we define
\begin{align}
 F_0^+ &= \rho  R_{ml}(\rho) \sqrt{\frac{2l+3}{2}} P_{l+1}(\cos\theta)\  ,\  {\tilde F}_0^+ = \rho  R_{ml}(\rho)\  ,\  F_J^+ = {\tilde F_J}^+\ \sqrt{\frac{2l+3}{2}}\ P_{l+1}(\cos\theta) \\
 F_0^- &= \rho  R_{ml}(\rho) \sqrt{\frac{2l-1}{2}}\ P_{l-1}(\cos\theta)\  ,\ {\tilde F}_0^- = \rho  R_{ml}(\rho)\  ,\  F_J^- = {\tilde F_J}^-\ \sqrt{\frac{2l-1}{2}}\ P_{l-1}(\cos\theta)
 \end{align}
then the generalization of (\ref{eq:I24}) may  be given in the form
\begin{align}
{\tilde F}_J^+\ &=\ \big(h_+\ +\ k_{m}^2\big)\ {\tilde F}_{J-1}^+\ \ \rightarrow \ \ {\tilde F}_J^+\ =\ \big(h_+\ +\ k_{m}^2\big)^J\ {\tilde F}_0^+ \label{eq:K1}\\
{\tilde F}_J^-\ &=\ \big(h_+\ +\ k_{m}^2\big)\ \ {\tilde F}_{J-1}^-\ \ \rightarrow \ \ {\tilde F}_J^-\ =\ \big(h_-\ +\ k_{m}^2\big)^J\ {\tilde F}_0^-  \label{eq:K2}
\end{align}
for positive $J$, where
\begin{align}
h_+ \  &=\ -\ \frac{\partial^2}{\partial \rho^2}\ +\ \frac{(l+1) (l+2)}{\rho^2}\ +\ 2 v_0 \\
h_-\  &=\ -\ \frac{\partial^2}{\partial \rho^2}\ +\ \frac{l(l-1)}{\rho^2}\ +\ 2 v_0 
\end{align}
For $l\ne 0$ the functions $ {\tilde F}_J^+$ and ${\tilde F}_J^-$ lead to separate sum rules. The above considerations lead us to consider the following sums valid for for any value of $l$ :
\begin{align}
S_J^+ &= 
\sum_n \big(k_m^2-k_n^2\big)^J  | \langle m,l| z |n,l+1 \rangle |^2 +\int_0^{\infty}dk \big(k_m^2 +k^2\big)^J | \langle m,l| z |k ,l+1\rangle |^2 =   \alpha^2 \langle {\tilde F}_K ^+| {\tilde F}_{J-K}^+\rangle  \label{eq:L1}\\
S_J^- &=  
\sum_n \big(k_m^2-k_n^2\big)^J  | \langle m,l| z |n,l-1 \rangle |^2 +\int_0^{\infty}dk \big(k_m^2 +k^2\big)^J | \langle m,l| z |k,l-1 \rangle |^2 = \beta^2\ \langle {\tilde F}_K^- | {\tilde F}_{J-K}^-\rangle \label{eq:L2}\\
S_J\ &=\ S_J^+\ +\ S_J^-  \label{eq:L3}
\end{align}
and the different values of $K$ for a fixed value of $J$ indicate the various equivalent forms for the sum rules. 
 
The functions ${\tilde F}_0^{\pm}$ and
\begin{align}
{\tilde F}_1^+\ &=\ \big(h^+ + k_m^2\big)\ \big(\rho R_{ml}\big)\ =\ 2 \Big(\frac{l+1}{\rho} - \frac{\partial}{\partial\rho}\Big) R_{ml}\ =\ -2 \rho^{l+1}\ \frac{\partial}{\partial\rho} \Big(\frac{R_{ml}}{\rho^{l+1}}\Big) \label{eq:K3}\\
{\tilde F}_1^-\ &=\ \big(h^- + k_m^2\big)\ \big(\rho R_{ml}\big)\ =\ 2 \Big(-\frac{l}{\rho} - \frac{\partial}{\partial\rho}\Big) R_{ml}\ =\ -\frac{2}{\rho^l}\ \frac{\partial}{\partial\rho}\Big(\rho^l R_{ml}\Big) \label{eq:K4} \\
{\tilde F}_2^+\ &=\ \big(h^+ + k_m^2\big)\ {\tilde F}_1^+\ =\ 4\ \frac{\partial v_0}{\partial\rho}\ R_{ml}  \label{eq:K5} \\
{\tilde F}_2^-\ &=\ \big(h^- + k_m^2\big)\ {\tilde F}_1^-\ =\ 4\ \frac{\partial v_0}{\partial\rho}\ R_{ml}  \label{eq:K6}\\
{\tilde F}_3^+\ &=\ \big(h^+ + k_m^2\big)\ {\tilde F}_2^+\ =\ 8\ \Big(\frac{l+1}{\rho^2}\ \frac{\partial v_0}{\partial\rho}\ -\ \frac{1}{2} \frac{\partial^3v_0}{\partial\rho^3}\ -\ \frac{\partial^2v_0}{\partial\rho^2}\ \frac{\partial}{\partial\rho}\Big)\ R_{ml} \label{eq:K7}\\
{\tilde F}_3^-\ &=\ \big(h^- + k_m^2\big)\ {\tilde F}_2^-\ =\ 8\ \Big(-\frac{l}{\rho^2}\ \frac{\partial v_0}{\partial\rho}\ -\ \frac{1}{2} \frac{\partial^3v_0}{\partial\rho^3}\ -\ \frac{\partial^2v_0}{\partial\rho^2}\ \frac{\partial}{\partial\rho}\Big)\ R_{ml}  \label{eq:K8}
\end{align}
can be used to examine the first few sum rules. For $J=0$ we find that
\begin{equation}
S_0^+ = \alpha^2\ \langle m,l | \rho^2 | m,l\rangle\ ,\ 
S_0^- = \beta^2\ \langle m,l | \rho^2 | m,l\rangle\ ,\ 
S_0   = \langle m,l | \rho^2\ \cos^2\theta | m,l\rangle = \langle z^2 \rangle
\end{equation}
For $J=1$ it may be shown that
\begin{align}
S_1^+\ &=\ \alpha^2\ \langle {\tilde F}_0^+\ |\ {\tilde  F}_1^+ \rangle\ =\ 2 \alpha^2\ \langle m,l | \Big(l+1\ -\ \rho\frac{\partial}{\partial\rho}\Big) | m,l \rangle\ =\ \frac{(l+1)^2}{2l+1} \\
S_1^-\ &=\ \beta^2\ \langle {\tilde F}_0^-\ |\ {\tilde  F}_1^- \rangle\ =\ 2 \beta^2\ \langle m,l | \Big(l+1\ -\ \rho\frac{\partial}{\partial\rho}\Big) | m,l \rangle\ =\ - \frac{l^2}{2l+1} \\
S_1\ &=\ 1
\end{align}
Proceeding along these lines the next 3 orders of sum rules can be given in the form
\begin{align}
 S_2^+ &= \alpha^2\ \langle {\tilde F}_0^+\ |\ {\tilde  F}_2^+ \rangle\ =\ 4\ \alpha^2\ \langle m,l\ |\ \rho \frac{\partial v_0}{\partial\rho}\ |\ m,l \rangle \\
 S_2^- &= \beta^2\ \langle {\tilde F}_0^-\ |\ {\tilde  F}_2^- \rangle\ =\ 4\ \beta^2\ \langle m,l \ |\ \rho \frac{\partial v_0}{\partial\rho}\ |\ m,l \rangle \\
S_2 &= 4\ \langle m,l\ |\ \rho \frac{\partial v_0}{\partial\rho}\ \cos^2\theta\ |\ m,l \rangle \ =\ 4\ \langle z\frac{\partial v_0}{\partial z}\rangle \\
 S_3^+ &= \alpha^2\ \langle {\tilde F}_1^+\ |\ {\tilde  F}_2^+ \rangle\ =\ 8\ \alpha^2\ \langle m,l\ |\ \frac{\partial v_0}{\partial\rho}\ \Big(\frac{l+1}{\rho}\ -\ \frac{\partial}{\partial\rho}\Big)\ |\ m,l\rangle \\
S_3^- &= \beta^2\ \langle {\tilde F}_1^-\ |\ {\tilde  F}_2^- \rangle\ =\ 8\ \beta^2\ \langle m,l\ |\ \frac{\partial v_0}{\partial\rho}\ \Big(-\frac{l}{\rho}\ -\ \frac{\partial}{\partial\rho}\Big)\ |\ m,l\rangle \\
S_3 &= 4\ \langle m,l\ |\ \frac{\partial v_0}{\partial\rho}\ \Big(\frac{\sin^2\theta}{\rho}\ -\ 2\cos^2\theta\ \frac{\partial}{\partial\rho}\Big) |\ m,l \rangle = 4 \langle\frac{\partial^2v_0}{\partial z^2}\rangle +4\langle\cos^2\theta\rangle \Big(\frac{\partial v_0}{\partial \rho}  R_{ml}^2\Big)|_{\rho=0}\\
S_4^+ &= \alpha^2\ \langle {\tilde F}_2^+\ |\ {\tilde  F}_2^+ \rangle\ =\ 16\ \alpha^2\ \langle m,l\ |\ \Big(\frac{\partial v_0}{\partial \rho} \Big)^2\ |\ m,l \rangle\\
S_4^- &= \beta^2\ \langle {\tilde F}_2^-\ |\ {\tilde  F}_2^- \rangle\ =\ 16\ \beta^2\ \langle m,l\ |\ \Big(\frac{\partial v_0}{\partial \rho} \Big)^2\ |\ m,l \rangle\\
S_4 &= 16\ \langle m,l\ |\ \Big(\frac{\partial v_0}{\partial \rho}\ \cos\theta \Big)^2\ |\ m,l \rangle\ =\ 16\ \langle \Big(\frac{\partial v_0}{\partial z}\Big)^2 \rangle
\end{align}
It may also be established that the fifth order sum rule may be given as 
\begin{align}
S_5\ &=\ \alpha^2\ \langle {\tilde F}_2^+\ |\ {\tilde  F}_3^+ \rangle\ +\ \beta^2\ \langle {\tilde F}_2^-\ |\ {\tilde  F}_3^- \rangle \notag\\
&=\ 16\ \langle\ \Big[\ \Big(\frac{\partial v_0}{\partial\rho}\Big)^2\ \frac{\sin^2\theta}{\rho^2}\ -\ \frac{\partial v_0}{\partial\rho}\ \cos^2\theta\ \Big(\frac{\partial^3 v_0}{\partial\rho^3}\ +\ 2 \frac{\partial^2 v_0}{\partial\rho^2}\ \frac{\partial}{\partial\rho}\Big)\ \Big]\ \rangle \notag\\
&=\ 16\ \langle\ \Big[\ \Big(\frac{\partial v_0}{\partial\rho}\Big)^2\ \frac{\sin^2\theta}{\rho^2}\ +\ \Big(\frac{\partial^2 v_0}{\partial\rho^2}\Big)^2\  \cos^2\theta\ \Big]\ \rangle\ +\ 16\ \langle\cos^2\theta\rangle\ \Big[ \frac{\partial v_0}{\partial\rho}\ \frac{\partial^2 v_0}{\partial\rho^2}\ R_{ml}^2\Big]|_{\rho=0} \notag\\
&=\ 16\ \langle \ \Big(\nabla \frac{\partial v_0}{\partial z}\Big)\ .\ \Big(\nabla \frac{\partial v_0}{\partial z}\Big)\ \rangle\ +\  16\ \big(\alpha^2 + \beta^2\big)\ \Big[ \frac{\partial v_0}{\partial\rho}\ \frac{\partial^2 v_0}{\partial\rho^2}\ R_{ml}^2\Big]|_{\rho=0} 
\end{align}
The sixth order sum rule may also be constructed using (\ref{eq:K7}) and (\ref{eq:K8}) and given in the form
\begin{align}
S_6 &=\ \alpha^2\ \langle {\tilde F}_3^+\ |\ {\tilde  F}_3^+ \rangle\ +\ \beta^2\ \langle {\tilde F}_3^-\ |\ {\tilde  F}_3^- \rangle \notag \\
&= 64 \langle \cos^2\theta \Big[ 2\Big(E_m -v_0 -\frac{l(l+1)}{2\rho^2}\Big)  \Big(\frac{\partial^2 v_0}{\partial\rho^2}\Big)^2 + \frac{l(l+1)}{\rho^4}  \Big(\frac{\partial v_0}{\partial\rho}\Big)^2 + \frac{3}{4} \Big(\frac{\partial^3 v_0}{\partial\rho^3}\Big)^2 + \frac{1}{2} \frac{\partial^2 v_0}{\partial\rho^2}  \frac{\partial^4 v_0}{\partial\rho^4}\Big]\rangle \notag\\
&\ + 32 \langle \sin^2\theta \Big[ \frac{1}{\rho^4} \Big( \rho \frac{\partial^2v_0}{\partial\rho^2} - \frac{\partial v_0}{\partial\rho}\Big)^2\Big]\rangle\ +\ \Delta \\
\Delta &=  64\Big[\langle \sin^2\theta \rangle\ \frac{\partial v_0}{\partial\rho}  \frac{\partial^2 v_0}{\partial\rho^2} \frac{R_{ml}^2}{\rho^2}\ +\ \langle \cos^2\theta \rangle \frac{\partial^3 v_0}{\partial\rho^3} \frac{\partial^2v_0}{\partial\rho^2} \frac{R_{ml}^2}{2}\ -\ \langle \cos^2\theta \rangle  \Big(\frac{\partial^2 v_0}{\partial\rho^2}\Big)^2 R_{ml} \frac{\partial R_{ml}}{\partial\rho}\Big]|_{\rho=0}
\end{align}
Higher members of the sequence ${\tilde F}_J^{\pm}$ may be found using (\ref{eq:K1}) -(\ref{eq:K8}) and higher order sum rules may be constructed.

The sum rule of order $J$ can be written in many equivalent forms by considering the overlap of functions  ${\tilde F}_K^{\pm}$ and ${\tilde F}_{J-K}^{\pm} $ with K taking different values. For example the sum rules for $J=2$ may be written in terms of the overlap of the functions $F_j^{\pm}$ corresponding to $j=0$ and $j=2$ or in terms of the overlap of functions corresponding to $j=1$ and $j=1$. It can be shown that this equivalence leads to the relation
\begin{equation}
\langle \rho\ \frac{\partial V_0}{\partial \rho} \rangle\ =\ \langle R_{ml} | \frac{l(l+1)}{\rho^2} | R_{ml} \rangle \ +\ \langle \frac{\partial R_{ml}}{\partial \rho} | \frac{\partial R_{ml}}{\partial \rho} \rangle\ =\ \langle 2(E_m -V_0) \rangle\ =\ 2\ \langle \frac{p^2}{2M} \rangle 
\end{equation}
which is the Virial theorem for the potential $V_0$.

For $J=3$ the two different ways of expressing the sum rule {\it viz}, from the overlap (1,2) or the overlap (0,3), leads to 2 equivalent expressions for $S_3$. For the case of $J=4$ the overlaps (2,2), (1,3) and (0,4) lead to  3 equivalent expressions for $S_4$. Such equivalences arise for all higher orders and are examples of Kramer's relations which is discussed in Messiah (1966) in the context of the Coulomb potential. Higher order sum rules lead to a hierarchy of equivalences which may be looked upon as generalizations of the Virial theorem. These issues will be discussed in a later section.

For sum rules with $J$ a negative integer the sums have squares of matrix elements weighted by negative powers of energy gaps. The naive generalization of (\ref{eq:I34}) and (\ref{eq:I35}) would be
\begin{align}
\big(h_+ + k_m^2\big)\  {\tilde G}_{J+1}^+\ &=\ {\tilde G}_J^+  \label{eq:K9}\\
\big(h_- + k_m^2\big)\  {\tilde G}_{J+1}^-\ &=\ {\tilde G}_J^-  \label{eq:K10}
\end{align}

However the inhomogeneous differential equations (\ref{eq:K9}) and (\ref{eq:K10}) must be solved with due consideration given to the possibility that the homogeneous part of the differential equations may have normalizable solutions satisfying bound state boundary conditions. If such solutions exist the normalized solutions to the homogeneous differential equations  denoted by
\begin{equation}
 \big(h_+ + k_m^2\big)\ {\tilde R}^+\ =\ 0\ ,\ \ \ \big(h_- + k_m^2\big)\ {\tilde R}^-\ =\ 0 
\end{equation}
give rise to the additional conditions on the solutions to the inhomogeneous differential equations: 
\begin{align}
\langle {\tilde R}^+ | \big(h_+ + k_m^2\big)\ {\tilde G}_{J+1}^+\rangle\ &=\ \langle {\tilde G}_{J+1}^+ | \big(h_+ + k_m^2\big)\ {\tilde R}^+\rangle^{\dagger}\ =\ 0 \\
\langle {\tilde R}^- | \big(h_- + k_m^2\big)\ {\tilde G}_{J+1}^-\rangle\ &=\ \langle {\tilde G}_{J+1}^- | \big(h_- + k_m^2\big)\ {\tilde R}^-\rangle^{\dagger}\ =\ 0 
\end{align}
Hence the normalized solutions to the homogeneous differential equations must be used to modify equations (\ref{eq:K9}) and (\ref{eq:K10}) to the form
\begin{align}
\big(h_+ + k_m^2\big)\  {\tilde G}_{J+1}^+\ &=\ {\bar G}_J^+\ =\ {\tilde G}_J^+ \ -\ \langle{\tilde G}_{J}^+| {\tilde R}^+\rangle\ {\tilde R}^+  \label{eq:K11}\\
\big(h_- + k_m^2\big)\  {\tilde G}_{J+1}^-\ &=\ {\bar G}_J^-\ ={\tilde G}_J^- \ -\ \langle{\tilde G}_{J}^-| {\tilde R}^-\rangle\ {\tilde R}^- \label{eq:K12}
\end{align}
so that the necessary conditions on the solutions to the inhomogeneous differential equations specified by (\ref{eq:K11}) and (\ref{eq:K12}) are automatically fulfilled by the new functions ${\bar G}$ :
\begin{equation}
\langle{\tilde R}^+ | {\bar G}_J^+\rangle\ =\ 0\ \ \ ,\ \ \ \langle{\tilde R}^- | {\bar G}_J^-\rangle\ =\ 0  
\end{equation}

The expressions for the sum rules for positive values of J given in (\ref{eq:L1}) and (\ref{eq:L3}) may now be extended to negative values of J with the identification ${\tilde F}_{-J}^{\pm} =\ {\bar G}_J^{\pm}$ and the negative order sum rules may be given in the form
\begin{align}
S_{-J}^+\ &=\  
\sum_{n\ne m} \frac{| \langle m,l| z |n,l+1 \rangle |^2}{\big(k_m^2-k_n^2\big)^J} +\int_0^{\infty}dk\frac{  | \langle m,l| z |k ,l+1\rangle |^2}{\big(k_m^2 +k^2\big)^J} =   \alpha^2 \langle {\bar G}_K ^+| {\bar G}_{J-K}^+\rangle \label{eq:L4}\\
S_{-J}^-\ &=\  
\sum_{n\ne m} \frac{| \langle m,l| z |n,l-1 \rangle |^2}{\big(k_m^2-k_n^2\big)^J} +\int_0^{\infty}dk\frac{  | \langle m,l| z |k ,l-1\rangle |^2}{\big(k_m^2 +k^2\big)^J} =   \beta^2 \langle {\bar G}_K ^-| {\bar G}_{J-K}^-\rangle  \label{eq:L5}\\
S_{-J}\ &=\ S_{-J}^+\ +\ S_{-J}^-  \label{eq:L6}
\end{align}
and the different values of $K$ for a fixed value of $J$ indicate the various equivalent forms for the sum rules. This construction is essentially the generalization of the procedure considered by Dalgarno and Lewis (1956). For $l\ne 0$ the functions $ {\bar G}_J^+$ and ${\bar G}_J^-$ lead to separate sum rules. 

We have shown that sums involving the squares of matrix elements connecting an initial state to final states in the discrete and continuum spectrum  by a dipole perturbation weighted by various positive powers of the energy differences may be evaluated by 3 different methods: 
(I) By evaluation of transition matrix elements to individual states and summation and integration. 
(II) By finding the functions ${\tilde F}_J^{\pm}$ and evaluating the overlap integrals of these functions. We have shown that there are many equivalent ways of choosing a pair of functions from the set of functions ${\tilde F }_J^{\pm}$ to evaluate the overlap integral relevant for  the sum rule of a particular order. 
(III) By evaluation of a single matrix element in the initial state as in (\ref{eq:I4}), (\ref{eq:I10}, (\ref{eq:I11}), (\ref{eq:I15}) and (\ref{eq:I17}). 
Sums with squares of matrix elements weighted by negative powers of energy differences may be evaluated using the methods (I) and (II) in the above list.

\section{Coulomb potential}

We now illustrate the procedure described above by examining the sum rules for the case of  an attractive Coulomb potential. For the Coulomb potential the right hand side of (\ref{eq:I4}), (\ref{eq:I10}), (\ref{eq:I11}), (\ref{eq:I16}) and (\ref{eq:I17}), which involve the expectation values of the derivatives of the Coulomb potential in the initial state $|m,l\rangle$, may be evaluated in closed analytic form and lead to  expressions for the first few positive order sum rules in terms of $m$ and the eigenvalue $\lambda=l(l+1)$ of ${\bf L}^2$ in the form
\begin{align}
S_0\ &=\ \langle m,l | \rho^2\ \cos^2\theta |m,l\rangle\ =\ \frac{m^2}{2}\ \big(5 m^2 +1 - 3\lambda\big)\  \frac{2\lambda -1}{4\lambda -3}\  \label{eq:L7}\\
S_1\ &=\ \langle m,l |m,l\rangle\ =\ 1 \label{eq:L8}\\
S_2\ &=\ 4\ \langle m,l |\ \frac{\cos^2\theta}{\rho}\ |m,l\rangle\ =\ \frac{4}{m^2}\ \frac{2\lambda -1}{4\lambda -3} \label{eq:L9}\\
S_3\ &=\ 4\ \langle m,l |\ \frac{-2\cos^2\theta + \sin^2\theta}{\rho^3}\ |m,l\rangle\ +\ 4\ \langle m,l| \cos^2\theta|m,l\rangle\ \psi_{ml}^2(0)\ \delta_{l,0} \notag\\
&=\  \frac{-16}{m^3} \ \frac{1}{4\lambda-3}\ \frac{1}{{\sqrt{4\lambda^2+1}}} \label{eq:L10}\\
S_4\ &=\ 16\ \langle m,l |\ \frac{\cos^2\theta}{\rho^4}\ |m,l\rangle\ =\ \Big(\frac{4}{m}\Big)^3\ \frac{3m^2 -\lambda}{m^2}\ \frac{2\lambda -1}{\lambda(4\lambda -3)^2}\ \frac{1}{{\sqrt{4\lambda^2+1}}},\ l\ge 1 \label{eq:L11}
\end{align}

\subsection{Sum rules for the 2S state of Hydrogen}

The normalized radial functions for the 2S and 2P states are
\begin{align}
R_{2,0}\ &=\ \frac{1}{{\sqrt 8}}\ (\rho^2 - 2 \rho)\ \exp \big(-\frac{\rho}{2}\big) \label{eq:L12}\\
R_{2,1}\ &=\ \frac{1}{{\sqrt 4!}}\ \rho^2\ \exp \big(-\frac{\rho}{2}\big) \label{eq:L13}
\end{align}
The 2S state is degenerate with the 2P state and a dipole perturbation can connect the 2S to all the P states including the 2P state. There is no need to consider ${\tilde F}_J^- $ as there are no states with angular momenta less than 0. The first few solutions for ${\tilde F}_J^+$ for positive values of J may be constructed in the form 
\begin{align}
{\tilde F}_0^+\ &=\ \rho\ R_{2,0}\ =\ \frac{1}{{\sqrt 8}}\ (\rho^3 - 2 \rho^2)\ \exp \big(-\frac{\rho}{2}\big) \label{eq:L14}\\
{\tilde F}_1^+\ &=\ \big(h_+ + \frac{1}{4}\big)\ {\tilde F}_0^+\ =\ \frac{1}{{\sqrt 8}}\ (\rho^2 - 4 \rho)\ \exp \big(-\frac{\rho}{2}\big) \label{eq:L15}\\
{\tilde F}_2^+\ &=\ \big(h_+ +\frac{1}{4}\big)\ {\tilde F}_1^+\ =\ {\sqrt 2}\ \big(1\ -\ \frac{2}{\rho}\big)\ \exp \big(-\frac{\rho}{2}\big) \label{eq:L16}
\end{align}
Since there is a normalizable solution to $\big(h_+ +\frac{1}{4}\big){\tilde R}^+ =0$ given by ${\tilde R}^+ = R_{2,1}$ given in (137) we find that
\begin{align}
{\bar G}_0^+\ &=\ {\tilde F}_0^+ - {\tilde R}^+ \langle {\tilde R}^+ | {\tilde F}_0^+ \rangle\\
\big(h^+ +\frac{1}{4}\big)\ {\tilde G}_1^+\ &=\ {\bar G}_0^+\ =\ \frac{1}{{\sqrt 8}}\ (\rho^3 - 5 \rho^2)\ \exp \big(-\frac{\rho}{2}\big)
\end{align}
which may be solved to give
\begin{equation}
{\tilde F}_{-1}^+\ =\ {\bar G}_1^+\ =\ \frac{1}{{\sqrt 8}}\ \big(\frac{\rho^4}{2}\ -\ 15 \rho^2\big)\ \exp \big(-\frac{\rho}{2}\big)
\end{equation}
so that the conditions $\langle R_{2,1} | {\tilde F}_J^+\rangle =0$ are satisfied when $J\ne0$. The overlap integrals of the ${\tilde F}_J^+$  in (\ref{eq:L1}) for $J\ge 0$ and the overlap integrals in (\ref{eq:L4}) for negative orders lead to the sum rules
\begin{equation}
S_0\ =\ 14\ ,\ S_1\ =\ 1\ ,\ S_2\ =\ \frac{1}{3}\ ,\ S_3\ =\ \frac{2}{3}\ ,\ S_{-1}\ =\ 30\ ,\ S_{-2}\ =\ 195  \label{eq:L17}
\end{equation}
The positive order sums are in agreement with the results given in (\ref{eq:L7})-(\ref{eq:L10}) for $m=2$ and $\lambda=0$. The numerical evaluation of the contribution of the first 2000 discrete states and the continuum states to the sum rules which are given below
\begin{align}
Order &\ \ \ Discrete\ 2S-nP && Continuum\ 2S-kP &&& Sum\ \ \ \   \notag      \\
 S_{0}       &\ \ \  13.176806..  &&\ \ \  0.823193..    &&&\ \ \  13.999999..\\
 S_{1}       &\ \ \  0.648907..  &&\ \ \  0.351092..    &&&\ \ \  0.999999..\\
 S_{2}       &\ \ \  0.104632..  &&\ \ \  0.228701..    &&&\ \ \  0.333333..\\
 S_{3}       &\ \ \  0.017622..  &&\ \ \  0.649044..    &&&\ \ \  0.666666.. \\
 S_{-1}      &\ \ \  27.70006..  &&\ \ \  2.29993..     &&&\ \ \ 29.99999..  \\
 S_{-2}      &\ \ \ 187.959..    &&\ \ \  7.04049..     &&&\ \ \ 194.9999..
\end{align}
verify that the sums tend towards the values stipulated by the sum rules in (\ref{eq:L17}). The continuum contribution to the sums with $J\ge 4$ diverge. Sum rules exist for all negative values of J and may be evaluated by solving inhomogeneous differential equations for ${\tilde F}_{-2}^+, {\tilde F}_{-3}^+$,....  and evaluating the overlap integrals of the appropriate functions.

\subsection{Sum rules for the 2P state of H}

The 2P state is degenerate with the 2S state and a dipole perturbation can connect the 2P state to all the S states and the higher lying D states. From (\ref{eq:K0}) it can be seen that $\alpha^2 = \frac{4}{15}$ and $\beta^2 = \frac {1}{3}$ for the excitations from the 2P state. Since there are no normalizable solutions to $\big(h_+ +\frac{1}{4}\big) {\tilde R}^+ =0$ the first few solutions for ${\tilde F}_J^+$ are 
\begin{align}
{\tilde F}_0^+\ &=\ \rho\ R_{2,1}\ =\ \frac{1}{{\sqrt 4!}}\ \rho^3 \ \exp \big(-\frac{\rho}{2}\big)\\
{\tilde F}_1^+\ &=\ \big(h_+ +\frac{1}{4}\big)\ {\tilde F}_0^+\ =\ \frac{1}{{\sqrt 4!}}\ \rho^2\ \exp \big(-\frac{\rho}{2}\big)\\
{\tilde F}_2^+\ &=\ \big(h_+ +\frac{1}{4}\big) {\tilde F}_1^+\ =\ \frac{4}{{\sqrt 4!}}\ \exp \big(-\frac{\rho}{2}\big)\\
{\tilde F}_{-1}^+\ &=\ \frac{1}{{\sqrt 4!}}\ \big(\frac{\rho^4}{2}\ +\ 3 \rho^3\big)\ \exp \big(-\frac{\rho}{2}\big)\\
{\tilde F}_{-2}^+\ &=\ \frac{1}{{\sqrt 4!}}\ \frac{1}{6}\ \big(\rho^5\ +\ 16 \rho^4\ +\ 96\ \rho^3\big)\ \exp \big(-\frac{\rho}{2}\big)
\end{align}
There is a normalizable solution to $\big(h_- + \frac{1}{4}\big) {\tilde R}^- =0$ which is ${\tilde R}^- = R_{2,0}$ given in (\ref{eq:L14}) and the first few solutions for ${\tilde F}_J^-$ can be constructed in the form
\begin{align}
{\tilde F}_0^-\ &=\ \rho R_{2,1} - R_{2,0}\ \langle R_{2,0} | R_{2,1}  \rangle \ =\ \frac{1}{{\sqrt 4!}}\ \big(\rho^3\ -\ 9\ \rho^2 +\ 18\ \rho\big) \ \exp \big(-\frac{\rho}{2}\big)\\
{\tilde F}_1^-\ &=\ \big(h_- +\frac{1}{4}\big)\ {\tilde F}_0^-\ =\ \frac{1}{{\sqrt 4!}}\ \big(\rho^2\ -\ 6\ \rho\big)\ \exp \big(-\frac{\rho}{2}\big)\\
{\tilde F}_2^-\ &=\ \big(h_- +\frac{1}{4}\big)\ {\tilde F}_1^-\ =\ \frac{4}{{\sqrt 4!}}\ \exp \big(-\frac{\rho}{2}\big)\\
{\tilde F}_{-1}^-\ &=\ \frac{1}{{\sqrt 4!}}\ \big(\frac{\rho^4}{2}\ -\ 3 \rho^3\ -\ 3\ \rho^2\ +\ 6\ \rho\big)\ \exp \big(-\frac{\rho}{2}\big)\\
{\tilde F}_{-2}^-\ &=\ \frac{1}{{\sqrt 4!}}\ \frac{1}{6}\ \big(\rho^5\ +\ \rho^4\ -\ 6\ \rho^3\ -456\ \rho^2\ +\ 912\ \rho\big)\ \exp \big(-\frac{\rho}{2}\big)
\end{align}
so that the conditions $\langle R_{2,0} | {\tilde F}_J^- \rangle =0$ are fulfilled. The overlap integrals of the ${\tilde F}_J^{\pm}$  in (\ref{eq:L1})-(\ref{eq:L3}) for $J\ge 0$ and the overlap integrals in (\ref{eq:L4})-(\ref{eq:L6}) for negative orders lead to the sum rules
\begin{align}
S_0^- &= 10\ ,\ S_1^- = -\frac{1}{3}\ ,\ S_2^- = \frac{1}{3}\ ,\ S_3^- =\ -\frac{2}{9}\ ,\ S_4^-\ =\ \frac{2}{9} ,\ S_{-1}^-\ =\ 2\ ,\ S_{-2}^-\ =\ 19 \label{eq:L18}\\
S_0^+ &=\ 8\ ,\ S_1^+\ =\ \frac{4}{3}\ ,\ S_2^+ = \frac{4}{15}\ ,\ S_3^+ =\ \frac{4}{45}\ ,\ S_4^+ = \frac{8}{45}\ ,\ S_{-1}^+ = 52\ ,\ S_{-2}^+ = 352 \label{eq:L19}\\
S_0\ &=\ 18\ ,\ S_1\ =\ 1\ ,\ S_2\ =\ \frac{3}{5}\ ,\ S_3\ =\ -\frac{2}{15}\ ,\ S_4\ =\ \frac{2}{5}\ ,\ S_{-1} = 54\ ,\ S_{-2}^- = 371  \label{eq:L20}
\end{align}
The positive order sums $S_J$ are in agreement with the results given in (\ref{eq:L7})-(\ref{eq:L11}) for $m=2$ and $\lambda=2$. The continuum contribution to the sum rules diverge for $J \ge 5$ but the sum rules converge for all negative orders. The numerical evaluation of the contribution of the first 2000 discrete states and the continuum states to the sum rules which are given below
\begin{align}
Order &\ \ \ Discrete\ 2P-nS && Continuum\ 2P-kS &&& Sum\ \ \ \   \notag      \\
 S_{0}^-       &\ \ \  9.93978..  &&\ \ \  0.06021..    &&&\ \ \  9.99999..\\
 S_{1}^-       &  -0.35677..  &&\ \ \  0.02344..    &&&\ \   -0.33333..\\
 S_{2}^-       &\ \ \  0.32166..  &&\ \ \  0.01167..    &&&\ \ \  0.33333..\\
 S_{3}^-       &  -0.23252..  &&\ \ \  0.01030..    &&&\ \   -0.22222.. \\
 S_{4}^-       &\ \ \  0.17586..  &&\ \ \  0.04636..    &&&\ \ \  0.22222..  \\
 S_{-1}^-      &\ \ \  1.82473..  &&\ \ \  0.17526..    &&&\ \ \  1.99999..  \\
 S_{-2}^-      &\ \ \  18.4514..  &&\ \ \  0.5485..     &&&\ \ \  18.9999..
\end{align}
\begin{align}
Order &\ \ \ Discrete\ 2P-nD && Continuum\ 2P-kD &&& Sum\ \ \ \   \notag     \\
 S_{0}^+       &\ \ \  7.38669..  &&\ \ \  0.61330..    &&&\ \ \  7.99999..\\
 S_{1}^+       &\ \ \  1.11382..  &&\ \ \  0.21951..    &&&\ \ \  1.33333..\\
 S_{2}^+       &\ \ \  0.17304..  &&\ \ \  0.09362..    &&&\ \ \  0.26666..\\
 S_{3}^+       &\ \ \  0.02790..  &&\ \ \  0.06098..    &&&\ \ \  0.08888..\\
 S_{4}^+       &\ \ \  0.00470..  &&\ \ \  0.17307..    &&&\ \ \  0.17777..\\
 S_{-1}^+      &\ \ \  50.1225..  &&\ \ \  1.87746..    &&&\ \ \  51.9999..\\
 S_{-2}^+      &\ \ \  345.927..  &&\ \ \  6.07274..     &&&\ \ \ 351.999..
\end{align}
verify that the sums tend towards the values stipulated by the sum rules in (\ref{eq:L18}) and (\ref{eq:L19}).


\subsection{Dipole sum rules for power law potentials}

We now consider the dipole sum rules for power law potentials $V_0=A r^\gamma$ expressed in terms of a dimensionless radial variable $\rho$ using an appropriate scaling length $a$ and a corresponding scaled  energy in the form
\begin{equation}
V_0\ =\ \epsilon\ v_0,\ \ v_0\ =\ \frac{\rho^\gamma}{\gamma} \ \ ,\ \ \epsilon\ =\ \frac{\hbar^2}{M a^2}, \ \ a^{\gamma+2}\ =\ \frac{\hbar^2}{M |A \gamma|}\ \ ,\ \ E_m\ =\ \epsilon\ \epsilon_m \label{eq:L21}
\end{equation}
and examine the dimensionless quantities 
\begin{equation}
S_J\ =\ \Big(\frac{2}{\epsilon}\Big)^J\ \Big[\sum_n\ \big(E_n - E_m)^J\  | \langle m|\ \rho \cos\theta |n \rangle |^2\ +\int_0^{\infty}dk\ \big(E_k - E_m\big)^J\ | \langle m|\ \rho\cos\theta\ |k \rangle |^2\Big] 
\end{equation}
Using the relations given in (\ref{eq:K50}, (\ref{eq:K0}), (\ref{eq:K51}), (\ref{eq:I4}), (\ref{eq:I10}), (\ref{eq:I11}), (\ref{eq:I15}) and (\ref{eq:I17}) it can be shown that the first few sum rules may be brought to the form
\begin{align}
S_0\ &=\ \frac{2l(l+1)-1}{4l(l+1)-3}\ \langle m |\ \rho^2\ | m\rangle \label{eq:L22}\\
S_1\ &=\ 1 \label{eq:L23}\\
S_2\ &=\ 4\ \frac{2l(l+1)-1}{4l(l+1)-3}\ \langle m |\ \rho\ \frac{\partial v_0}{\partial \rho} \ | m\rangle\\
S_3\ &=\ 4\ \langle m |\ \Big[\frac{\partial^2v_0}{\partial\rho^2}\ \cos^2\theta\ +\frac{1}{\rho}\ \frac{\partial v_0}{\partial\rho}\ \sin^2\theta \Big]\ |m\rangle \notag\\
S_4\ &=\ 16\ \frac{2l(l+1)-1}{4l(l+1)-3}\ \langle m| \Big(\frac{\partial v_0}{\partial \rho}\Big)^2\ |m\rangle
\end{align}
The Virial theorem applied to power law potentials 
\begin{equation}
\langle\rho\ \frac{\partial v_0}{\partial\rho} \rangle\ =\ \gamma \ \langle v_0\rangle\ =\ 2\ \langle(\epsilon_m - v_0)\rangle\ \rightarrow\ \langle v_0\rangle\ =\ \frac{2}{\gamma+2}\ \epsilon_m
\end{equation}
may be used to express $S_2$ in the form
\begin{equation}
S_2\ =\ 4\ \frac{2\gamma}{\gamma+2}\ \epsilon_{m} \ \frac{2l(l+1)-1}{4l(l+1)-3} 
\end{equation}

Using (\ref{eq:K0})  and (\ref{eq:L21}) the expressions for $S_3$ and $S_4$ can be brought to the form
\begin{align}
S_3\ &=\ 4 \ \Big[\frac{2l(l+1)-1}{4l(l+1)-3}\ (\gamma -2)\  +\ 1\Big]\ \langle m|\ \rho^{\gamma-2}\ |m\rangle \\
S_4\ &=\ 16 \ \frac{2l(l+1)-1}{4l(l+1)-3}\ \langle m|\ \rho^{2\gamma-2}\ |m\rangle
\end{align}

The potential $V_0 = \epsilon \log \rho$ does not belong to the power law category but $S_0$ and $S_1$ are still given by (\ref{eq:L22}) and (\ref{eq:L23}) and 
\begin{align}
S_2\ &=\  4 \ \frac{2l(l+1)-1}{4l(l+1)-3} \\
S_3\ &=\ 4 \ \langle m |\ \frac{\sin^2\theta - \cos^2\theta}{\rho^2}\ | m\rangle\  =\  \frac{4}{3-4l(l+1)}\ \langle m |\ \frac{1}{\rho^2}\ | m \rangle \\
S_4\ &=\ 16 \ \frac{2l(l+1)-1}{4l(l+1)-3}\ \langle m |\ \frac{1}{\rho^2}\ | m \rangle\ =\ 4 \big[1-2l(l+1)]\ S_3
\end{align}
It is evident that $S_3$ and $S_4$ are simply related by a scaling factor. The Virial theorem for a logarithmic potential leads to the relation
\begin{equation}
 \langle m | \log\rho | m\rangle =\epsilon_m - \frac{1}{2} 
\end{equation}

\section{Equivalence relations}

It was noted earlier that the sum rules may be represented in various equivalent forms. Here we consider the question in detail. In general it is possible to establish that for functions obeying relations of the form of (\ref{eq:K1}) and (\ref{eq:K2})
\begin{align}
\langle F_j | F_{k+1}\rangle\ &=\ \int_0^{\infty} F_j\ \big(h_{\pm} - E_{m}\big)\ F_k\ d\rho\ =\ \langle F_{j+1} | F_k\rangle\ -\ W\big(F_j,F_k\big)|_0^\infty  \\
W\big(F_j,F_k\big)\ &=\ F_j\ \frac{\partial F_k}{\partial\rho}\ -\ F_k\ \frac{\partial F_j}{\partial\rho}
\end{align}
For exponentially decaying functions the Wronskian evaluated at $\rho=\infty$ vanishes and if the overlap integrals exist it is possible to establish that the two overlaps are equal only if W evaluated at $\rho=0$ vanishes. But if the $W(\rho=0)$ exists and has a non-vanishing value $\Delta$ then in general 
\begin{equation}
\langle F_j | F_{k+1}\rangle\ =\ \langle F_{j+1} | F_{k}\rangle\ +\ \Delta
\end{equation}
By evaluating ${\tilde F}_3^{\pm}$ using (\ref{eq:K7}) and (\ref{eq:K8}) it can be established that
\begin{align}
\langle {\tilde F}_0^{\pm} | {\tilde F}_3^{\pm}\rangle\ &=\ \langle {\tilde F}_1^{\pm} | {\tilde F}_2^{\pm}\rangle\ +\ \Delta^{\pm} \\
\Delta^{\pm}\ &=\ W\big({\tilde F}_0^{\pm},{\tilde F}_2^{\pm}\big)|_0\ =\ 4 \Big[ R_{ml}^2\ \Big(\rho\ \frac{\partial^2v_0}{\partial\rho^2}\ -\ \frac{\partial v_0}{\partial\rho}\Big) \Big] |_0
\end{align}
If $ v_0(0)\ \rightarrow\ b\rho^{\gamma}|_0$ and $R_{ml}(0)\ \rightarrow\ C_l \ \rho^{l+1}|_0$ then
\begin{equation}
\Delta^{\pm}\ =\ 4 bC_l^2\ \gamma(\gamma-2)\ \ \hbox{if}\ \ l= -\frac{\gamma+1}{2}\ ,\ \Delta\ =\ 0\ \ \hbox{otherwise}
\end{equation}  

For the Coulomb potential using
\begin{equation}
b\ =\ -1\ ,\ \gamma\ =\ -1,\ C_l\ =\ \frac{2^{l+1}}{m^{l+2}} \frac{1}{(2l+1)!}\ \frac{{\sqrt{ (m+l)!}}}{{\sqrt{ (m-l-1)!}}}
\end{equation}
it can be shown that 
\begin{equation}
l=0\ \rightarrow\ \Delta^{\pm}\ =\ -\frac{48}{m^3} \ \ ,\ \ l\ne 0\ \rightarrow\ \Delta^{\pm}\ =\ 0
\end{equation} 
We can thus conclude that for Hydrogen atom the sum rule of order 3 can be given in the form
\begin{align}
l=0\ &\rightarrow\ S_3^{+}\ =\ \frac{1}{3}\ \langle {\tilde F}_1^{+} | {\tilde F}_2^{+}\rangle\ =\ \frac{1}{3}\ \langle {\tilde F}_0^{+} | {\tilde F}_3^{+} \rangle\ +\ \frac{16}{m^3} \label{eq:M1}\\
l\ne 0\ &\rightarrow\ S_3^{+}\ =\ \alpha^2\ \langle {\tilde F}_1^{+} | {\tilde F}_2^{+}\rangle\ =\ \alpha^2\ \langle {\tilde F}_0^{+} | {\tilde F}_3^{+}\rangle \label{eq:M2}\\
l\ne 0\ &\rightarrow\ S_3^{-}\ =\ \beta^2\ \langle {\tilde F}_1^{-} | {\tilde F}_2^{-}\rangle\ =\ \beta^2\ \langle {\tilde F}_0^{-} | {\tilde F}_3^{-}\rangle \label{eq:M3}
\end{align}
We can check this result using the example of the 2S state of Hydrogen for which using (\ref{eq:L14})-(\ref{eq:L16}) and
\begin{equation}
{\tilde F}_3^+\ =\ \Big(h_+\ +\ \frac{1}{4}\Big) {\tilde F}_2^+\ =\ {\sqrt 8}\ \Big( -\ \frac{1}{\rho}\ +\ \frac{4}{\rho^2}\Big)\ \exp \big(-\frac{\rho}{2}\big) \label{eq:M4}
\end{equation}
it can be shown that
\begin{equation}
\langle {\tilde F}_1^+ | {\tilde F}_2^+\rangle\ =\ 2\ ,\ \ \langle {\tilde F}_0^+ | {\tilde F}_3^+\rangle\ =\ -4
\end{equation}
which satisfy (\ref{eq:M1}) with $m=2$. We have used the example of Hydrogen to illustrate the subtlety required in establishing the equivalence of different expressions of the sum rules. In the case of Hydrogen we can use (\ref{eq:I16}) and the result that the Laplacian acting on the Coulomb potential yields a delta function to easily establish that for the S states
\begin{equation}
S_3\ =\ \frac{4}{3}\ \Big(R_{m,0}(0)\Big)^2\ =\ \frac{16}{3m^3}
\end{equation}
which for $m=2$ gives $S_3=\frac{2}{3}$ in agreement with (\ref{eq:M1}) and (\ref{eq:M4}). For all potentials $v_0$ which are less singular than the Coulomb potential $S_3$ is finite and may be expressed in terms of $v_0$ in the equivalent forms
\begin{equation}
S_3\ =\ 4\langle \frac{\partial^2v_0}{\partial z^2}\rangle\ =\ 4 \langle\Big[\frac{\partial v_0}{\partial\rho} \frac{\sin^2\theta}{\rho}\ +\ \cos^2\theta\ \frac{\partial^2v_0}{\partial\rho^2}\Big] \rangle\ =\ 4 \langle \frac{\partial v_0}{\partial\rho}\ \Big[\frac{\sin^2\theta}{\rho}\ -\ 2\cos^2\theta\ \frac{\partial}{\partial\rho}\Big] \rangle 
\end{equation}

The procedure discussed above can be extended to study equivalent expressions for the higher order sum rules. For the fourth order we can establish that
\begin{align}
\langle {\tilde F}_2^{\pm} | {\tilde F}_2^{\pm}\rangle &= \langle {\tilde F}_1^{\pm} | {\tilde F}_3^{\pm}\rangle\ -\ W\big({\tilde F}_1^{\pm},{\tilde F}_2^{\pm}\big)|_0 \\
W\big({\tilde F}_1^{+},{\tilde F}_2^{+}\big)|_0 &= 8\Big[\frac{\partial^2v_o}{\partial\rho^2} \Big(\frac{l+1}{\rho} R_{ml} - \frac{\partial R_{ml}}{\partial\rho}\Big) R_{ml} \notag\\
 &+ \frac{\partial v_0}{\partial\rho}\Big(\frac{l+1}{\rho^2} R_{ml}^2 - \Big(\frac{\partial R_{ml}}{\partial\rho}\Big)^2 + R_{ml} \frac{\partial^2R_{ml}}{\partial\rho^2}\Big)\Big]|_0\ \rightarrow 0\\
W\big({\tilde F}_1^{-},{\tilde F}_2^{-}\big)|_0 &= -8\Big[\frac{\partial^2v_o}{\partial\rho^2} \Big(\frac{l}{\rho} R_{ml} + \frac{\partial R_{ml}}{\partial\rho}\Big) R_{ml}
 + \frac{\partial v_0}{\partial\rho}\Big(\frac{l}{\rho^2} R_{ml}^2 + \Big(\frac{\partial R_{ml}}{\partial\rho}\Big)^2 - R_{ml} \frac{\partial^2R_{ml}}{\partial\rho^2}\Big)\Big]|_0 \notag\\ 
&\rightarrow -8(2l+1)\Big[\Big(\frac{\partial^2v_o}{\partial\rho^2} \frac{1}{\rho} + \frac{\partial v_0}{\partial\rho} \frac{1}{\rho^2}\Big) R_{ml}^2\Big]|_0
\end{align}
The Wronskian may be evaluated for the Coulomb potential using known solutions to show that it vanishes for $l\ge2$, is divergent for $l=0$ and for $l=1$ it has a finite value given by
\begin{equation}
 W\big({\tilde F}_1^{-},{\tilde F}_2^{-}\big)|_0 \ =\ 24\ \frac{2^2}{3^2}\ \frac{m^2-1}{m^5}\ =\ \frac{32}{3}\ \frac{(m^2-1)}{m^5}
 \end{equation}
This result may be checked for the 2P state of Hydrogen for which all the integrals may be explicitly carried out yielding
\begin{equation}
\langle {\tilde F}_2^{+} | {\tilde F}_2^{+}\rangle = \langle {\tilde F}_2^{-} | {\tilde F}_2^{-}\rangle\ =\ \langle {\tilde F}_1^{+} | {\tilde F}_3^{+}\rangle\ =\ \frac{2}{3}\ \ ,\ \ 
\langle {\tilde F}_1^{-} | {\tilde F}_3^{-}\rangle = \frac{5}{3}\ \ ,\ \
W\big({\tilde F}_1^{-},{\tilde F}_2^{-}\big)|_0  = 1
\end{equation}
in agreement with the general results. 

Another equivalence to consider in the fourth order is
\begin{equation}
\langle {\tilde F}_1^{\pm} | {\tilde F}_3^{\pm}\rangle\ =\ \langle {\tilde F}_0^{\pm} | {\tilde F}_4^{\pm}\rangle\ -\ W\big({\tilde F}_0^{\pm},{\tilde F}_3^{\pm}\big)|_0 
\end{equation}
The Wronskians can be evaluated for the Coulomb potential by considering the limiting values of ${\tilde F}_J^{\pm}$ as $\rho\rightarrow 0$ and it can be shown that it vanishes for $l\ge2$ and yields values $\Delta^{\pm}$ for $l=1$. It can be shown that $\Delta^-=0$ and $\Delta^+$ is finite.

Thus the fourth order sum rule for the Coulomb potential can be given as
\begin{align}
l\ge 2\  \rightarrow\ S_4^{+} &= \alpha^2\ \langle {\tilde F}_2^{+} | {\tilde F}_2^{+}\rangle =  \alpha^2\ \langle {\tilde F}_1^{+} | {\tilde F}_3^{+}\rangle = \alpha^2\ \langle {\tilde F}_0^{+} | {\tilde F}_4^{+}\rangle  \label{eq:M5}\\
l=1\ \rightarrow\ S_4^{+} &= \alpha^2\ \langle {\tilde F}_2^{+} | {\tilde F}_2^{+}\rangle = \alpha^2\ \langle {\tilde F}_1^{+} | {\tilde F}_3^{+}\rangle = \alpha^2\ \Big(\langle {\tilde F}_0^{+} | {\tilde F}_4^{+}\rangle  + \frac{160(m^2-1)}{3m^5}\ \Big) \label{eq:M6}\\
l\ge2\ \rightarrow\ S_4^{-} &= \beta^2\ \langle {\tilde F}_2^{-} | {\tilde F}_2^{-}\rangle =  \beta^2\ \langle {\tilde F}_1^{-} | {\tilde F}_3^{-}\rangle =\  \beta^2\ \langle {\tilde F}_0^{-} | {\tilde F}_4^{-}\rangle  \label{eq:M7}\\
l=1\ \rightarrow\ S_4^{-}\ &=\ \beta^2\ \langle {\tilde F}_2^{-} | {\tilde F}_2^{-}\rangle = \beta^2\ \Big(\ \langle {\tilde F}_1^{-} | {\tilde F}_3^{-}\rangle \ -\  \frac{32(m^2-1)}{3m^5}\ \Big) \notag\\
& = \beta^2\ \Big(\ \langle {\tilde F}_0^{-} | {\tilde F}_4^{-}\rangle  -  \frac{32(m^2-1)}{3m^5}\ \Big) \label{eq:M8}
\end{align}
For the Coulomb potential $S_4$ is finite only if $l\ge 1$. For all potentials $v_0$ which are less singular than the Coulomb potential $S_4$ is finite for $l\ge1$ and may be expressed in terms of $v_0$ in the equivalent forms
\begin{align}
&\frac{S_4}{16} = \langle \Big(\frac{\partial v_0}{\partial \rho}\Big)^2\rangle  = \langle \Big[  \Big(-k_m^2 -2 v_0 - \frac{l(l+1}{\rho^2}\Big) \frac{\partial^2v_0}{\partial \rho^2} + \frac{l(l+1)}{\rho^3} \frac{\partial v_0}{\partial \rho} + \frac{1}{4} \frac{\partial^4 v_0}{\partial \rho^4} \Big]\rangle  \label{eq:M9}\\
=& \langle \rho \Big[(2v_0+k_m^2) \frac{\partial^3 v_0}{\partial \rho^3} + \frac{l(l+1)}{\rho^2}\Big(\frac{\partial^3v_0}{\partial \rho^3} - \frac{1}{\rho} \frac{\partial^2v_0}{\partial \rho^2} + \frac{1}{\rho^2} \frac{\partial v_0}{\partial \rho}\Big) + \frac{\partial v_0}{\partial \rho} \frac{\partial^2v_0}{\partial \rho^2}- \frac{1}{4} \frac{\partial^5v_0}{\partial \rho^5} - \frac{1}{2\rho} \frac{\partial^4v_0}{\partial \rho^4} \Big] \rangle \label{eq:M10}
\end{align}    
For the Coulomb potential if $l=1$ there is an additional term arising from the non-vanishing contribution at $\rho=0$ of terms arising from integration by parts which can be calculated. But for $l\ge2$, (\ref{eq:M6}) and (\ref{eq:M7}) are valid and gives rise to the relation  
\begin{align}
\langle \frac{1}{\rho^4}\rangle\ &=\ \frac{2}{m^2} \langle\frac{1}{\rho^3}\rangle\ -\ 4\  \langle \frac{1}{\rho^4}\rangle\ +\ 3(l-1)(l+2)\ \langle \frac{1}{\rho^5}\rangle \\
&=\ \frac{6}{m^2}\ \langle \frac{1}{\rho^3}\rangle\ -\ 14\ \langle \frac{1}{\rho^4}\rangle\ +\ 9(l-1)(l+2)\ \langle \frac{1}{\rho^5}\rangle \\
\rightarrow\ 5\ \langle\frac{1}{\rho^4}\rangle\ &=\ \frac{2}{m^2} \langle\frac{1}{\rho^3}\rangle\ +\ 3\ (l-1)(l+2)\ \langle \frac{1}{\rho^5}\rangle \label{eq:M11} 
\end{align}
which is an example of Kramers relation ( Kramers 1926) for the Coulomb potential (Messiah 1966)
\begin{equation}
\frac{J+1}{m^2}\ \langle \rho^J\rangle\ -\ (2J+1)\ \langle \rho^{J-1}\rangle\ +\ \frac{J}{4}\  (2l+1+J)(2l+1-J)\ \langle \rho^{J-2}\rangle\ =\ 0, \ \ J\ge-2 l
\end{equation}
for the specific case $J=-3$.

The fifth, sixth and higher order sum rules may be given in many equivalent forms by considering overlaps of ${\tilde F}_J^{\pm}$ functions. To understand the relation between the different forms of the sum rules we next examine some properties of expectation values in Quantum Mechanics arising from the Schr{\"{o}}dinger equation.

\section{Generalization of Kramers relations}

We now show how the results derived in Quigg and Rossner (1979) and Messiah may be generalized. The Schr{\"{o}}dinger equation for the radial eigenfunction given in eq.(\ref{eq:I41}) may be written in the form
\begin{equation}
\frac{\partial^2 R_{ml}}{\partial\rho^2}\ =\ Q\ R_{ml}\ ,\ \ Q\ =\ \Big(k_m^2\ +\ 2v_0\ +\ \frac{l(l+1)}{\rho^2}\Big) \label{eq:N1}
\end{equation}
Let $f$ be a function of $\rho$ and a dot denote derivative with respect to $\rho$. Multiplication of the two sides of the Schr{\"{o}}dinger  equation by $2f{\dot R}_{ml}$  and integration in the domain $[0,\infty]$ yields the expressions
\begin{align}
\int_0^\infty f \frac{\partial{\dot R}_{ml}^2}{\partial\rho} d\rho &=  -f {\dot R}_{ml}^2|_0  - \int_0^\infty \Big({\dot f} {\dot R}_{ml}\Big) {\dot R}_{ml} d\rho \notag\\
&= \Big[-f {\dot R}_{ml}^2 + {\dot f}  {\dot R}_{ml} R_{ml}\Big]|_0\ +\int_0^\infty R_{ml} \Big(\frac{\partial^2f}{\partial\rho^2} {\dot R}_{ml} + {\dot f}\frac{\partial^2R_{ml}}{\partial\rho^2}\Big) d\rho \notag \\
&= \Big[-f {\dot R}_{ml}^2 + {\dot f} {\dot R}_{ml} R_{ml} - \frac{\partial^2f}{\partial\rho^2} \frac{R_{ml}^2}{2}\Big]|_0 +  \int_0^\infty \Big[-\frac{1}{2} \frac{\partial^3f}{\partial\rho^3} R_{ml}^2  +  {\dot f} Q R_{ml}^2\Big] d\rho \label{eq:N2}
\end{align}
and
\begin{align}
\int_0^\infty 2 f {\dot R}_{ml}\ Q\ R_{ml} d\rho\ &=\ -\Big[f Q R_{ml}^2\Big]|_0 - \int_0^\infty R_{ml}^2 \frac{\partial}{\partial\rho}\big(Qf\big)\ d\rho \notag\\
&= -\Big[f R_{ml} \frac{\partial^2R_{ml}}{\partial\rho^2}\Big]|_0 - \int_0^\infty R_{ml}^2 \frac{\partial}{\partial\rho}\big(Qf\big)\ d\rho \label{eq:N3}
\end{align}
where we have assumed that $f$ is such that the contributions from the boundary terms arising from the integrations by parts vanish at $\rho\rightarrow \infty$. The limiting values
\begin{equation}
f|_0\ \rightarrow\  b\ \rho^q\ ,\ \ R_{ml}|_0\ \rightarrow\ C_l \rho^{l+1}
\end{equation}
and the equality of eq.(\ref{eq:N2}) and (\ref{eq:N3}) may now be used to establish the relation
\begin{equation}
\frac{1}{2} \langle ml|\frac{\partial^3f}{\partial\rho^3}| ml\rangle\ -\ \langle ml|\frac{1}{f} \frac{\partial}{\partial\rho}\big(Qf^2\big) |ml\rangle\ +\ b\ C_l^2 \ (2l+1)^2\ \delta_{q,-2l}\ =\ 0\ , \ (q+2l) \ge 0 \label{eq:N4}
\end{equation}
The explicit form of $Q$ in eq.(\ref{eq:N1}) may be used to bring eq.(\ref{eq:N4}) to the form
\begin{equation}
{\bullet} -\frac{1}{4} \langle\frac{\partial^3f}{\partial\rho^3}\rangle + k_m^2 \langle\frac{\partial f}{\partial\rho}\rangle + \langle \frac{1}{f} \frac{\partial}{\partial\rho}\big(v_0f^2\big)\rangle + l(l+1) \langle\frac{1}{\rho}\frac{\partial}{\partial\rho}\Big(\frac{f}{\rho}\Big)\rangle \ =\ \frac{b}{2}\  C_l^2\ (2l+1)^2\ \delta_{q,-2l} \label{eq:N5}
\end{equation}
which reduces to (\ref{eq:M11}) if $f=\rho^{-2},\ v_0=-\rho^{-1}$ and $l\ge 2$. So (\ref{eq:N5}) is a generalization of the relations considered by Kramers (1926), Messiah (1966), and Quigg and Rossner (1979). 

We now consider some specific choices of $f$ which lead to interesting results.

1. If $f=b$ is a constant then we obtain the relation
\begin{equation}
V_{eff}\ =\ v_0\ +\ \frac{l(l+1)}{2\rho^2}\ ,\ \ \langle \frac{\partial V_{eff}}{\partial\rho}\rangle\ =\ \frac{C_l^2}{2}\ \delta_{0,l}  \label{eq:N6}
\end{equation}
which shows that the expectation value of the effective force is zero for non-zero values of $l$. For $l=0$ the expectation value of the force is only proportional to the square of the eigenfunction at $\rho=0$. The probability density $|\psi_{m,0}(0)|^2$ is of interest, for example, in the leptonic decays of massive neutral vector mesons $V^0$ which are $^3S_1$ bound states of a quark and an anti-quark. The decay width (Quigg and Rossner 1979) is given by
\begin{equation}
\Gamma\big(V^0\rightarrow l^+l^-\big)\ =\ 16\pi\ \frac{\hbar^3}{c}\ \Big(\frac{\alpha_e\ e_q}{M_V}\Big)^2\ |\Psi(0)|^2\ =\ 4\ \frac{c\hbar}{a} \Big(\frac{\hbar\ \alpha_e\ e_q}{M_V\ c\ a}\Big)^2\ C_0^2 \label{eq:N7}
\end{equation}
where $e_q$ is the charge of the quark in units of the electron charge, $c$ is the velocity of light and $M_V$ is the mass of the vector meson. It is evident from  (\ref{eq:N6}) and (\ref{eq:N7}) that the decay width is directly proportional to the expectation value of the force in a spherically symmetric eigenstate of the vector meson.

2. If $f=b\rho$ then we get
\begin{equation}
\langle \rho\ \frac{\partial v_0}{\partial\rho} \rangle\ =\ \langle \big(-k_m^2 - 2v_0\big)\rangle \ =\ 2\ \langle \big(E_m -v_0\big)\rangle\ =\ 2\ \langle {\hbox {Kinetic Energy}} \rangle
\end{equation}
which is the Virial theorem.

3. If $f=b\rho^2$ then 
\begin{equation}
\langle\ \Big[\ 2\rho\ (k_m^2\ +\ 2v_0)\ +\ \frac{l(l+1)}{\rho}\ +\ \rho^2\ \frac{\partial v_0}{\partial \rho}\ \Big]\ \rangle\ =\ 0
\end{equation}

4. If $f=b\rho^3$ then 
\begin{equation}
\langle\ \Big[\ 3\rho^2\ (k_m^2\ +\ 2v_0)\ +\ \rho^3\ \frac{\partial v_0}{\partial \rho}\ \Big]\ \rangle\ =\ - \frac{1}{2}\ (2l-1)(2l+3)
\end{equation}

5. If $f= b R_{ml}^2$ then
\begin{equation}
\langle \frac{\partial^3 f}{\partial \rho^3}\rangle\ =\ 0\ \ \rightarrow\ \ \int_0^\infty R_{ml}^2\ \frac{\partial^3 R_{ml}^2}{\partial \rho^3}\ d\rho\ =\ 0
\end{equation}
which is a statement about any bound state probability distribution.

6. If $f = v_0$ and $f(0)\rightarrow b\rho^q|_0$ then
\begin{equation}
\langle\ \Big[ -\frac{1}{4} \frac{\partial^3 v_0}{\partial \rho^3} +\ \big(3 v_0\ +\ k_m^2\big) \frac{\partial v_0}{\partial\rho}\ +\ \frac{l(l+1)}{\rho}\ \frac{\partial}{\partial\rho}\Big(\frac{v_0}{\rho}\Big)\Big]\ \rangle\ =\ 
\frac{1}{2}\ b\ C_l^2\ (2l+1)^2\ \delta_{q,-2l}
\end{equation}

7. If $f = \frac{\partial v_0}{\partial\rho}$ and $f(0)\rightarrow b\rho^q|_0$ then
\begin{equation}
\langle\ \Big[ -\frac{1}{4} \frac{\partial^4 v_0}{\partial \rho^4} + \Big(\frac{\partial v_0}{\partial\rho}\Big)^2 + \big(2 v_0 + k_m^2\big) \frac{\partial^2 v_0}{\partial\rho^2} + \frac{l(l+1)}{\rho^2}\ \frac{\partial}{\partial\rho}\Big(\frac{1}{\rho} \frac{\partial v_0}{\partial\rho}\Big)\Big]\ \rangle\ =\ 
\frac{b}{2}\ C_l^2\ (2l+1)^2\ \delta_{q,-2l} \label{eq:N8}
\end{equation}
which can be used to give the fourth order sum rule in the form
\begin{equation}
\frac{S_4}{16} = 
\langle \Big(\frac{\partial v_0}{\partial \rho}\Big)^2\rangle  =  \langle \Big[- \Big(k_m^2 + 2 v_0 + \frac{l(l+1}{\rho^2}\Big) \frac{\partial^2 v_0}{\partial \rho^2} + \frac{l(l+1)}{\rho^3} \frac{\partial v_0}{\partial \rho} + \frac{1}{4} \frac{\partial^4v_0}{\partial \rho^4} \Big]\rangle + \frac{b}{2}\ C_l^2\ (2l+1)^2\   \delta_{q,-2l} \label{eq:N9}
\end{equation}
in agreement with eq.(\ref{eq:M9}). Similarly the choice $f = \rho \frac{\partial^2 v_0}{\partial\rho^2}$ in eq. (\ref{eq:N5}) may be used to show that
\begin{align}
\langle \Big(\frac{\partial v_0}{\partial \rho}\Big)^2\rangle = &\langle \rho \Big[(k_m^2+2v_0) \frac{\partial^3 v_0}{\partial \rho^3}  + \frac{\partial v_0}{\partial \rho} \frac{\partial^2v_0}{\partial \rho^2} +  \frac{l(l+1)}{\rho^2}\Big(\frac{\partial^3v_0}{\partial \rho^3} - \frac{1}{\rho} \frac{\partial^2v_0}{\partial \rho^2} + \frac{1}{\rho^2} \frac{\partial v_0}{\partial \rho}\Big)  \notag\\
&- \frac{1}{4} \frac{\partial^5v_0}{\partial \rho^5} - \frac{1}{2\rho} \frac{\partial^4v_0}{\partial \rho^4} \Big] \rangle\ +\ \frac{b}{2}\ C_l^2\  (2l+1)^3\ \delta_{q,-2l}  \label{N9}
\end{align}
which is consistent with  (\ref{eq:M9}) and (\ref{eq:M10}). It is clear that the equivalence expressed in (\ref{eq:N5}) is responsible for the various different ways of expressing the sum rules.

All the above results can be verified using the known expectation values for the Coulomb potential (Pauling and Wilson 1935) :
\begin{align}
\langle\rho^2\rangle\ &=\ \frac{m^4}{2}\ \Big(5\ +\ \frac{\big(1-3l(l+1)\big)}{m^2}\ \Big)\ ,\ \ \langle\rho\rangle\ =\ \frac{m^2}{2}\ \Big( 3\ -\ \frac{l(l+1)}{m^2}\ \Big) \\
\langle\rho^{-1}\rangle\ &=\ \frac{1}{m^2}\ ,\ \ \langle\rho^{-2}\rangle\ =\ \frac{1}{m^3}\ \frac{2}{2l+1}\\
\langle\rho^{-3}\rangle\ &=\ \frac{1}{m^3}\ \frac{2}{l(l+1)(2l+1)}\ ,\ l\ge 1 \label{eq:N10}\\
\langle\rho^{-4}\rangle\ &=\ \frac{1}{m^3}\ \Big(3\ -\ \frac{l(l+1)}{m^2}\Big)\ \frac{4}{l(l+1)(2l+3)(2l+1)(2l-1)}\ ,\ l\ge 1 \label{eq:N11}
\end{align}
For example, for the $l=1$ states (\ref{eq:N8}) and (\ref{eq:N9}) lead to the equality
\begin{equation}
\langle \frac{1}{\rho^4}\rangle\ =\ \frac{1}{m^2} \langle \frac{2}{\rho^3}\rangle - \langle\frac{4}{\rho^4}\rangle + \frac{9}{2}\ \frac{2^2(m^2-1)}{3^2 m^5}\ =\ \frac{1}{m^2}\langle\frac{6}{\rho^3}\rangle - \langle\frac{12}{\rho^4}\rangle -\langle\frac{2}{\rho^4}\rangle  +  \frac{27}{2}\ \frac{2^2(m^2-1)}{3^2 m^5}
\end{equation} 
which may be verified using (\ref{eq:N10}) and (\ref{eq:N11}).

In this report we have shown that the solutions to the radial Schr{\"{o}}dinger equation for effective one-dimensional systems may be used to construct a hierarchy of sum rules satisfied by certain matrix elements thereby generalizing the well known sum rules satisfied by oscillator strengths.

\section{References}

[1] Jackiw, R. 1967 {\it {Phys. Rev.}} {\bf 157} 1220.

[2] Bethe, H.A. 1964 {\it {Intermediate Quantum Mechanics}} (W.A.Benjamin,Inc., New York)

[3] Kramers, H. A. 1926 {\it{Zeitschrift f{\"{u}}r Physik}} {\bf 39} 836

[4] Foot, C.J. 2013 {\it{Atomic Physics}} (Oxford University Press) 125

[5] Schiff, L. 1968 {\it {Quantum Mechanics}} (McGraw-Hill) 266

[6] Dalgarno, A. and Lewis, J.T. 1956 {\it {Proc. Roy. Soc.}} (London) {\bf A233} 70

[7] Messiah, A. 1966 {\it {Quantum Mechanics}} (North-Holland) 431

[8] Quigg, C. and Rosner, J.L. 1979 {\it {Physics Reports}} {\bf 56} 186

[9] Pauling, L. and E.B  Wilson 1935 {\it{Quantum Mechanics}} (NcGraw-Hill) 145

\end{document}